\documentclass[a4paper,fleqn]{cas-sc}

\usepackage{hyperref}

\usepackage[round]{natbib}

\usepackage{graphicx}
\usepackage{xspace}	
\usepackage{graphicx}
\usepackage{lscape}
\usepackage{threeparttable}
\usepackage{amsmath}
\usepackage[utf8]{inputenc}
\usepackage{longtable}
\usepackage{graphicx}
\usepackage{multirow}
\usepackage{rotating}

\newcommand{\apj}{The Astrophysical Journal}
\newcommand{\pasp}{Publications of the Astronomical Society of the Pacific}

\newcommand{\mnras}{Monthly Notices of the Royal Astronomical Society}

\newcommand{\physrep}{Physics Reports}
\newcommand{\nat}{Nature}
\newcommand{\apjl}{The Astrophysical Journal Letters}

\newcommand{\aap}{Astronomy \& Astrophysics}

\newcommand{\aj}{Astronomical Journal}
\newcommand{\apjs}{Astrophysical Journal Supplement}
\newcommand{\ssr}{Space Science Reviews}
\newcommand{\araa}{Annual Review of Astron and Astrophys}

\newcommand{\AstroSat}{{\em AstroSat}\xspace}
\newcommand{\fermi}{{\em Fermi}\xspace}
\newcommand{\kw}{{\em Konus}-Wind\xspace}

\newcommand{\swiftT}{{T$_{\rm 0}$}\xspace}
\newcommand{\fermiT}{{T$_{\rm 0}$}\xspace}
\newcommand{\keV}{{\rm keV}\xspace}

\newcommand{\swift}{{\em Swift}\xspace}
\newcommand{\tninty}{{$T_{\rm 90}$}\xspace}
\newcommand{\mvts}{{$t_{\rm mvts}$}\xspace}

\newcommand{\Ep}{$E_{\rm p}$\xspace}
\newcommand{\sw}[1]{\texttt{#1}}

\def\tsc#1{\csdef{#1}{\textsc{\lowercase{#1}}\xspace}}
\tsc{WGM}
\tsc{QE}
\tsc{EP}
\tsc{PMS}
\tsc{BEC}
\tsc{DE}

\begin{document}
\let\WriteBookmarks\relax
\def\floatpagepagefraction{1}
\def\textpagefraction{.001}

\shorttitle{GRB-SNe: Magnetar origin?}

\shortauthors{Amit Kumar et~al.}

\title [mode = title]{Tale of GRB~171010A/SN~2017htp and GRB~171205A/SN~2017iuk: Magnetar origin?}                      
\tnotemark[1,2]

\author[1,2]{Amit Kumar}[type=author,
                        auid=000,bioid=1,
                        orcid=0000-0001-7511-2910]

\cormark[1]


\affiliation[1]{organization={Aryabhatta Research Institute of Observational Sciences},
    addressline={Manora Peak}, 
    city={Nainital, Uttarakhand},
    postcode={263002}, 
    country={India; amitkundu515@gmail.com, shashi@aries.res.in, rahulbhu.c157@gmail.com}}

\affiliation[2]{organization={School of Studies in Physics and Astrophysics, Pandit Ravishankar Shukla University},
    addressline={Raipur}, 
    city={Chhattisgarh},
    postcode={492010}, 
    country={India}}

\author[1]{Shashi B. Pandey}
\author[1,3]{Rahul Gupta}
\affiliation[3]{organization={Department of Physics, Deen Dayal Upadhyaya Gorakhpur University},
    addressline={Gorakhpur}, 
    city={Uttar Pradesh},
    postcode={273009}, 
    country={India}}
    
\author[1,3]{Amar Aryan}

\author[1]{Amit K. Ror}

\author[1]{Saurabh Sharma}

\author[2]{Nameeta Brahme}

\cortext[cor1]{Corresponding author}

\begin{abstract}
We present late-time optical follow-up observations of GRB 171010A/SN 2017htp ($z$ = 0.33) and low-luminosity GRB 171205A/SN 2017iuk ($z$ = 0.037) acquired using the 4K$\times$4K CCD Imager mounted at the 3.6m Devasthal Optical Telescope (3.6m DOT) along with the prompt emission data analysis of these two interesting bursts. The prompt characteristics (other than brightness) such as spectral hardness, \tninty, and minimum variability time-scale are comparable for both the bursts. The isotropic $X$-ray and kinetic energies of the plateau phase of GRB~171205A are found to be less than the maximum energy budget of magnetars, supporting magnetar as a central engine powering source. The new optical data of SN~2017htp and SN~2017iuk presented here, along with published ones, indicate that SN~2017htp is one of the brightest and SN 21017iuk is among the faintest GRB associated SNe (GRB-SNe). Semi-analytical light-curve modelling of SN~2017htp, SN~2017iuk and only known GRB associated superluminous supernova (SLSN 2011kl) are performed using the {\tt MINIM} code. The model with a spin-down millisecond magnetar as a central engine powering source nicely reproduced the bolometric light curves of all three GRB-SNe mentioned above. The magnetar central engines for SN~2017htp, SN~2017iuk, and SLSN~2011kl exhibit values of initial spin periods higher and magnetic fields closer to those observed for long GRBs and H-deficient SLSNe. Detection of these rare events at such late epochs also demonstrates the capabilities of the 3.6m DOT for deep imaging considering longitudinal advantage in the era of time-domain astronomy.
\end{abstract}

\begin{keywords}
GRB-SNe connection: magnetar \sep Individual: GRB~171010A/SN~2017htp \sep GRB~171205A/SN~2017iuk \sep GRB~111209A/SLSN~2011kl.
\end{keywords}

\maketitle

\section{Introduction}
\label{Intro}
Gamma-Ray Bursts (GRBs) are short and highly energetic flashes of radiation occurring at cosmological distances and exhibit non-thermal spectra peaking in the $\gamma$-ray band \citep{Meegan1992, Band1993, Kumar2015, Peer2015}. Due to their high intrinsic luminosities, GRBs are observable up to very high redshifts (z $\sim$8-10). Thus, these are among the potential candidates for increasing our understanding of high-energy physical mechanisms \citep{Meszaros2013}, probing the properties of the primordial universe \citep{Fiore2001, Fynbo2007} and measuring the cosmological parameters \citep{Amati2013, Moresco2022}. Along with many observed prompt emission properties, bursts with $T_{\rm 90}$ duration \footnote{the period over which 90\% of the entire background-subtracted counts are observed} less than 2s are termed as short/hard GRBs (sGRBs), while those last for more than 2s are designated as long/soft GRBs (lGRBs), see \cite{Kouveliotou1993, Zhang2012a}. Recently, a few ultra-long GRBs (ulGRBs) have also been detected, lasting much longer (a few hundred seconds to hours) in $\gamma$-rays \citep{Boer2015, Perna2018, Dagoneau2020}, e.g., ulGRB 111209A was active for around $\sim$25000s \citep{Levan2014}.

Some of the lGRBs, including ultra-long and low-luminosity GRBs (llGRBs, $L_{\gamma,iso}$ $<$ $10^{48.5}$ erg s$^{-1}$; \citealt{Cano2017}), have exhibited correlations with H-deficient stripped-envelope supernovae (SESNe; \citealt{Wang1998, Nomoto2006, Woosley2006, Modjaz2011, Hjorth2012}), mostly with Ic broad-line SNe \citep[Ic-BL, see][for a review]{Cano2017}. SESNe comprise a small fraction of the known population of SNe. \cite{Li2011} proposed the rate of SESNe $\approx$16\% w.r.t. population of all known SNe, estimated by examining a set of 726 SNe observed utilising the Lick Observatory Supernova Search (LOSS; \citealt{Filippenko2001}). In addition, via investigating 117 known SNe followed by the Supernova Diversity and Rate Evolution \citep[SUDARE;][]{Botticella2013}, \cite{Cappellaro2015} also claimed a rate of SESNe $\approx$15\%. Whereas, among SESNe, only a small fraction of SNe~Ic-BL (nearly 7\%; \citealt{Guetta2007}) and an even smaller fraction of GRB associated SNe cases (GRB/SNe; $\sim$4\%) are observed \citep{Fryer2007}. At the same time, the rates of normal GRBs and llGRBs w.r.t. SESNe appear $\sim$0.4--3\% and 1--9\%, respectively \citep{Guetta2007}. Due to this sparsity, up to now, only a few dozen GRB/SNe within a redshift range of 0.00866 (GRB 980425/SN 1998bw; \citealt{Galama1998}) to 1.0585 (GRB 000911/SN; \citealt{Lazzati2001}) are detected. Although, some of the GRB/SNe show 3$-$5 orders of lower luminosities than average, making them challenging to detect at high redshifts \citep{Schulze2014}. The connection observed between ulGRB~111209A and superluminous (SL) SN 2011kl (@$z$ = 0.677) has opened a new window to explore whether SLSNe are also associated with lGRBs \citep{Greiner2015, Kann2019}. In addition, recent shreds of evidence of possible links between SLSNe and Fast Radio Bursts (FRBs) have devised further opportunities to look for the properties of associated magnetars as central engines in more detail \citep[][and references therein]{Eftekhari2019, Mondal2020}. Light-curve modelling of such events can add value to a better understanding of the possible powering mechanisms of such peculiar and rare transients.

In the case of SNe associated with GRBs (GRB-SNe), nearly 2$-$8 $M_\odot$ of material ejects, of which $\approx$0.1$-$0.5 $M_\odot$ is supposed to be radioactive $^{56}Ni$ \citep{Bloom1998, Berger2011, Cano2017}; whereas synthesised $^{56}Ni$ mass (M$_{Ni}$) can be higher for the events like ulGRB 111209A/SLSN 2011kl \citep{Kann2019}. The physical processes that power underlying SN arise via thermal heating from radioactive material trapped in the ejecta. Although, apart from the conventional SN power-source models (e.g., radioactive decay of $^{56}$Ni and circumstellar interaction), a central engine based powering source (a black hole/collapsar model; \citealt{Woosley1993} or a millisecond magnetar; \citealt{Usov1992}) can also explain observed properties of GRB-SNe \citep{MacFadyen1999, Wheeler2000, Woosley2006, Cano2016, Dessart2017, Obergaulinger2020, Roy2021, Zou2021}. The millisecond magnetars with initial spin periods of $\sim$1-10 ms and magnetic fields of $\sim$10$^{14-15}$ gauss are believed to induce relativistic Poynting-flux jets \citep{Bucciantini2008}, relevant mechanism to explain such energetic stellar explosions. As suggested by \cite{Duncan1992}, a rotating magnetar with a spin period of one millisecond, the mass of 1.4 M$_\odot$, and a radius of 10 kilometres reserves a rotational energy of $\sim$2.2 $\times$ 10$^{53}$ erg. A misaligned magnetar model for magnetar thermalisation and jet formation can also be used to explain possible connections between lGRBs and SNe \citep{Margalit2018}.

GRB 171010A was categorised as a lGRB at $z$ = 0.33 \citep{Frederiks2017, Omodei2017, Kankare2017}. \cite{Chand2019} studied the prompt radiation mechanism of the burst using spectro-polarimetric observations carried by \AstroSat. Optical photometric and spectroscopic studies of GRB 171010A/SN 2017htp and its host analysis are performed in detail by \cite{Melandri2019}. SN 2017htp started to emerge after $\sim$3d and peaked at $\sim$13d since associated GRB detection. The light-curve modelling of SN~2017htp was performed by considering radioactive decay of $^{56}$Ni as a primary powering source and to estimate the M$_{Ni}$, ejecta mass (M$_{ej}$) and the kinetic energy (E$_{k}$).

On the other hand, GRB 171205A was discovered as one of the most nearby lGRBs with $z$ = 0.037 \citep{DElia2017,DElia2018, Izzo2017a}. GRB 171205A was also categorized as a llGRB ($L_{\gamma,iso}$ $\sim$3 $\times$ $10^{47}$ erg s$^{-1}$; \citealt{DElia2018}) with a comparatively lower black body (BB) temperature and significant late time re-brightening \citep{Postigo2017, DElia2018, Wang2018, Izzo2019, Suzuki2019}. Based on the spectral modelling of GRB 171205A, \cite{Izzo2019} suggested that at the early times ($<$3d post burst), the energy injected by the GRB jets into the SN envelope created a cocoon-like structure. However, at late phases ($>$3d post burst), the light produced by the associated SN dominated and outshined the emission from the cocoon. Very late-time radio observations (around 1000d post burst; GHz to sub-GHz) of GRB 171205A were presented by \cite{Maity2021}, and they also suggested a hot cocoon scenario for GRB 171205A/SN 2017iuk. Similar to the case of GRB 171010A, the light-curve modelling of SN~2017iuk was performed using the radioactive decay model and estimated the M$_{Ni}$, M$_{ej}$, and E$_{k}$ \citep{Izzo2019}.

This article presents the prompt emission data analysis of GRB~171010A and GRB~171205A, late-time optical observations and the semi-analytical light-curve modelling of SN~2017htp and SN~2017iuk (along with SLSN~2011kl) to investigate the underlying physical mechanisms behind such rare events and to constrain some of the physical parameters. The optical observations of the above-discussed GRB-SNe are acquired using the first light instrument called 4K$\times$4K CCD Imager mounted at the axial port of the recently commissioned 3.6m Devasthal Optical Telescope \citep[3.6m DOT;][]{Pandey2018, Kumar2022}. The longitudinal advantage of India and specifically the 3.6m DOT at Nainital is crucial, having a significant temporal gap with its site in the middle of the longitude belt of 180$^\circ$ between Eastern Australia (160$^\circ$ E) and the Canary Islands (20$^\circ$ W), vital for transient studies at optical-NIR frequencies \citep{Kumar2018, Pandey2018, Sagar2019, Kumar2020b,Kumar2021a}.

The paper is organised as follows. Observations, data reduction, photometric calibrations, and the light curves production of GRB~171010A/SN~2017htp and GRB~171205A/SN~2017iuk are discussed in Section~\ref{sec:observations}. Results obtained from the high energy aspects and light-curve modelling on the optical data are presented in Section~\ref{sec:results}. We summarized our results in Section~\ref{sec:conclusion}. Throughout the analysis, H$_o$ = 70 km s$^{-1}$, $\Omega_m$ = 0.27, and $\Omega_\lambda$ = 0.73 are adopted to estimate the distances, and the magnitudes are used in the AB system.

\section{Observations, data processing and light curve production}
\label{sec:observations}

This section discusses the prompt analysis, optical observations, data reduction, photometric calibrations and light-curve production of GRB~171010A/SN~2017htp and GRB~171205A/SN~2017iuk.

\subsection{Prompt Gamma-ray/$X$-ray properties}

\begin{figure*}[!t]
\centering
\includegraphics[scale=0.32]{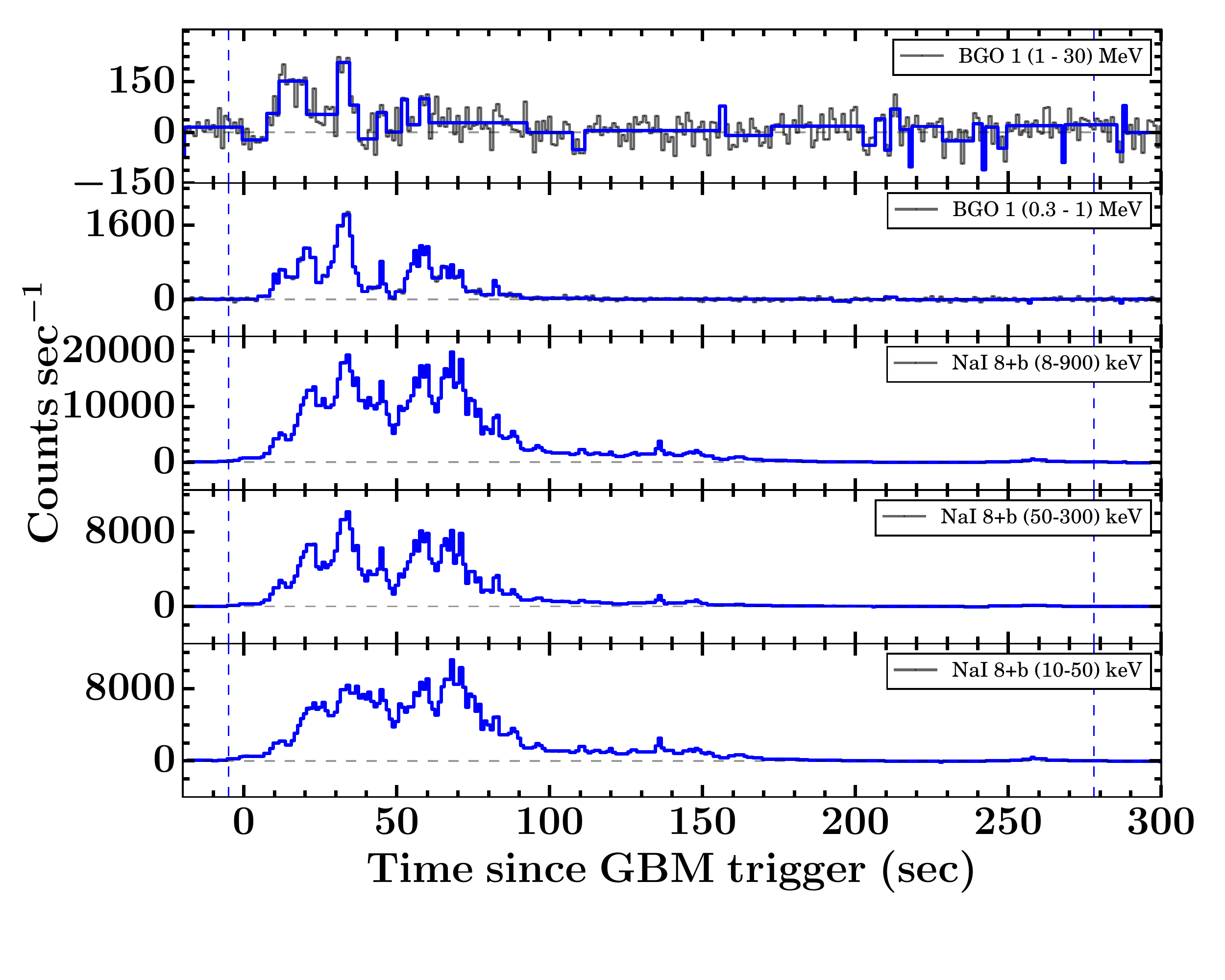}
\includegraphics[scale=0.32]{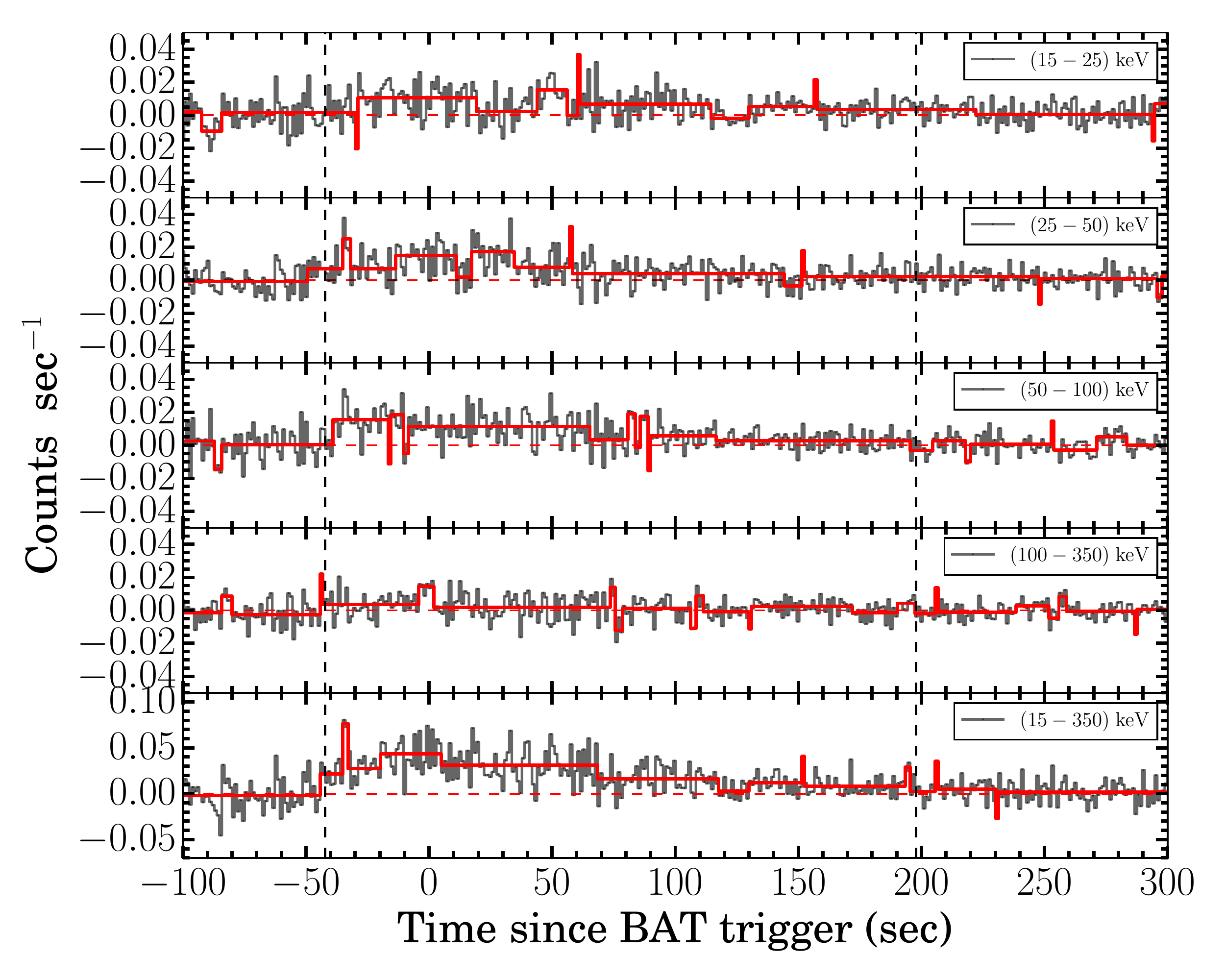}
\caption{{\it Prompt light curves$-$} Left panel: the \fermi GBM energy-resolved and background-subtracted light curves of GRB~171010A. The Bayesian blocks (blue) are overplotted to track the change in the light curve. The blue vertical dashed lines mark the time slice used for the GBM time-averaged spectral analysis. Right panel: the BAT energy-resolved mask-weighted light curves of GRB 171205A. The Bayesian block light curve for respective energy channels is shown in red colour. The black vertical dashed lines mark the time slice used for the BAT time-averaged spectral analysis.}\label{batlc}
\end{figure*}

GRB 171010A was discovered by the \fermi Large Area Telescope (LAT; \citealt{Atwood2009}) and Gamma-ray Burst Monitor (GBM; \citealt{Meegan2009}) on 2017-10-10 UT 19:00:50.58 (JD = 2458037.292; \citealt{Omodei2017, Poolakkil2017}) at $\alpha$ = 04$^{h}$ 26$^{m}$ 19$^s$.46 and $\delta$ = $-10^\circ$ 27$^{'}$ 45$^{''}$.9 (J2000). We obtained the \fermi GBM data of GRB 171010A from the \fermi GBM Burst Catalog\footnote{\url{https://heasarc.gsfc.nasa.gov/W3Browse/fermi/fermigbrst.html}} and performed the temporal and spectral data analysis following the methodology presented in \cite{2022MNRAS.511.1694G, 2022arXiv220507790C}. For the temporal analysis of the GBM data, we have used \sw{RMFIT} software and utilized the two brightest sodium iodide detectors (NaI-8 and NaI-b) and the brightest bismuth germanate detector (BGO-1). The \fermi GBM light curve of GRB 171010A consists of two different emission phases. The first main bright emission phase (from \fermiT-5 to \fermiT+205s) comprises multiple merging pulses followed by a very faint phase (from \fermiT+246 to \fermiT+278s). There is a quiescent gap of $\sim$41s between both the emission phases. The left panel of Figure \ref{batlc} shows the count-rate \fermi GBM light curve of GRB 171010A in five different energy ranges. The Bayesian blocks are overplotted on each light curve to track the change in counts. For the spectral analysis (to fit the time-integrated spectra), we have used the Multi-Mission Maximum Likelihood framework \citep[\sw{3ML}\footnote{\url{https://threeml.readthedocs.io/en/latest/}}]{2015arXiv150708343V} tool and many empirical models (\sw{Band}, \sw{Band+Blackbody}, \sw{Cutoff-power law}, and \sw{Bkn2pow}).

GRB 171205A was discovered by the Burst Alert Telescope (BAT; \citealt{Barthelmy2005}) on 2017-12-05 UT 07:20:43 (JD = 2458092.80605) at J200 coordinates: $\alpha$ = 11$^{h}$ 09$^{m}$ 39$^s$.46 and $\delta$ = $-12^\circ$ 35$^{'}$ 08$^{''}$.5 with an uncertainty box of three arcmins \citep{DElia2017}. We obtained the \swift BAT data from the \swift archive page of GRB 171205A\footnote{\url{https://www.swift.ac.uk/archive/selectseq.php?source=obs&tid=794972}} and performed the temporal and spectral data analysis following the methodology presented in \cite{2021MNRAS.505.4086G}. The mask-weighted \swift BAT prompt emission light curve of GRB 171205A presents some weak emissions with multiple overlapping peaked structures with a \tninty duration (in BAT 15-350 \keV energy range) of 189.4 $\pm$ 35s \citep{2017GCN.22184....1B}. The right panel of Figure \ref{batlc} presents the BAT light curve of GRB 171205A in different energy channels along with the Bayesian block. For the spectral modelling of GRB 171205A, we have used \sw{3ML} software (Bayesian fitting). We used different models such as power-law, cutoff-power law, and BB to fit the spectrum and utilised Bayesian Information Criteria \citep[BIC;][]{Kass:1995} to find the best fit model. The best fit spectral parameters results of both the bursts are discussed in Section \ref{promptspectra}.

\begin{figure*}[ht!]
\centering
\includegraphics[angle=0,scale=0.2205]{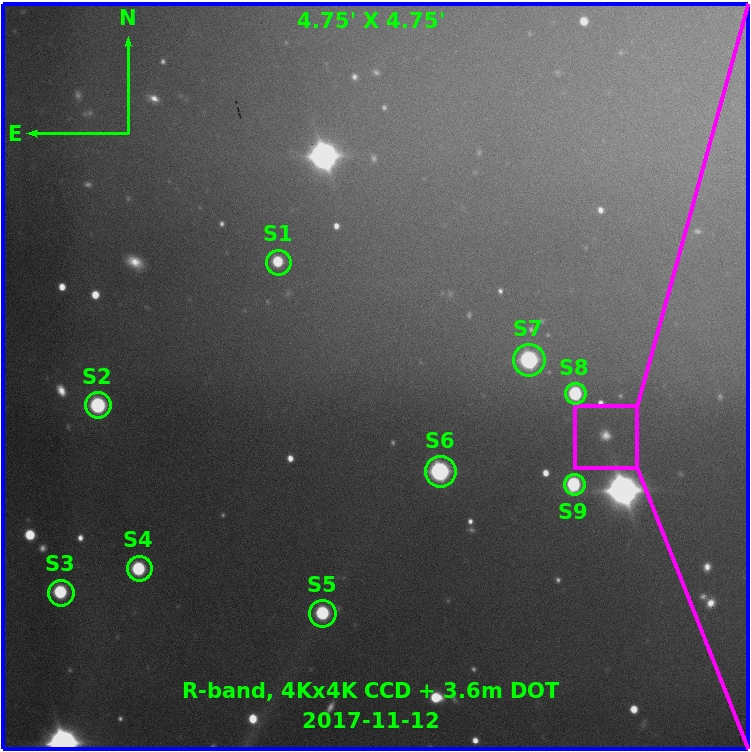}
\includegraphics[angle=0,scale=0.207]{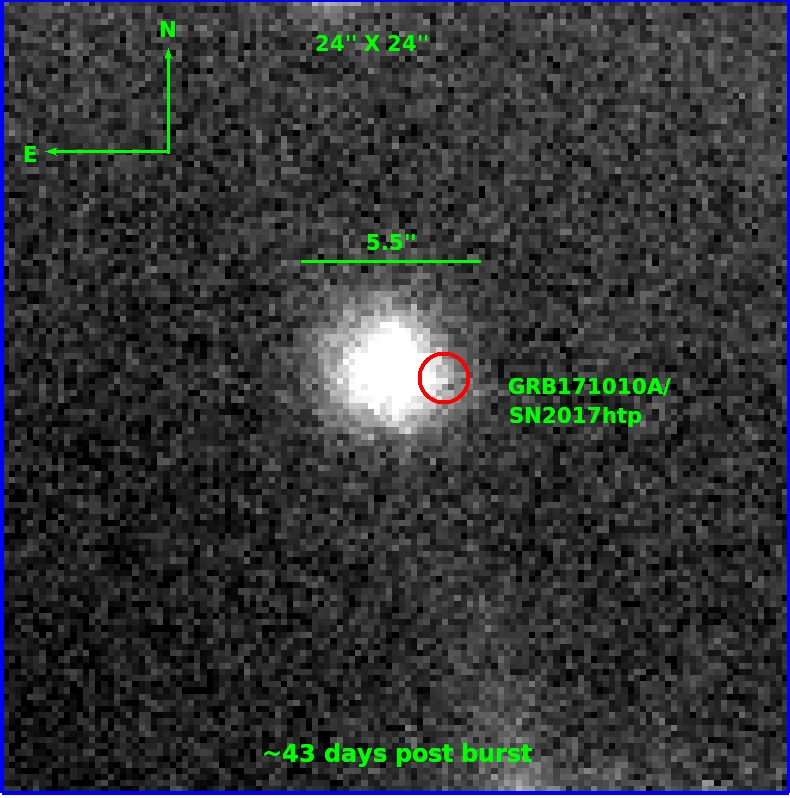}
\caption{Left panel: a section of stacked $R$-band image (8$\times$300s; 4.75$^{'}$ $\times$ 4.75$^{'}$) of GRB~171010A/SN~2017htp field observed on 2017-11-22 using the 4K$\times$4K CCD Imager at the 3.6m DOT. Secondary standard stars (S1-–S9) are encircled; those were used for calibration. Right panel: a 24$^{''}$ $\times$ 24$^{''}$ section of the GRB field is shown to highlight the GRB~171010A/SN~2017htp embedded in the host galaxy.}
\label{fig:finding_171010A}
\end{figure*}

\begin{figure*}[ht!]
\centering
\includegraphics[angle=0,scale=0.235]{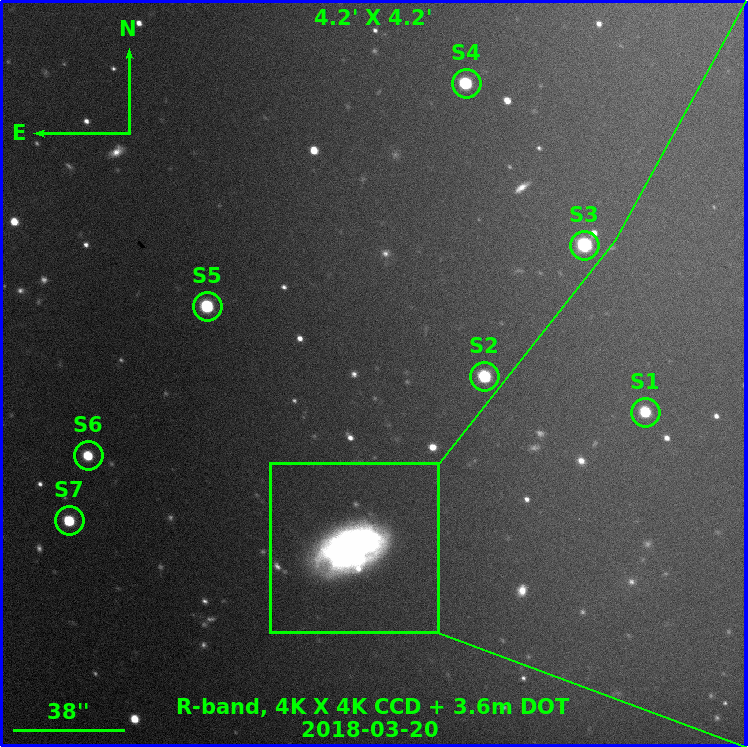}
\includegraphics[angle=0,scale=0.243]{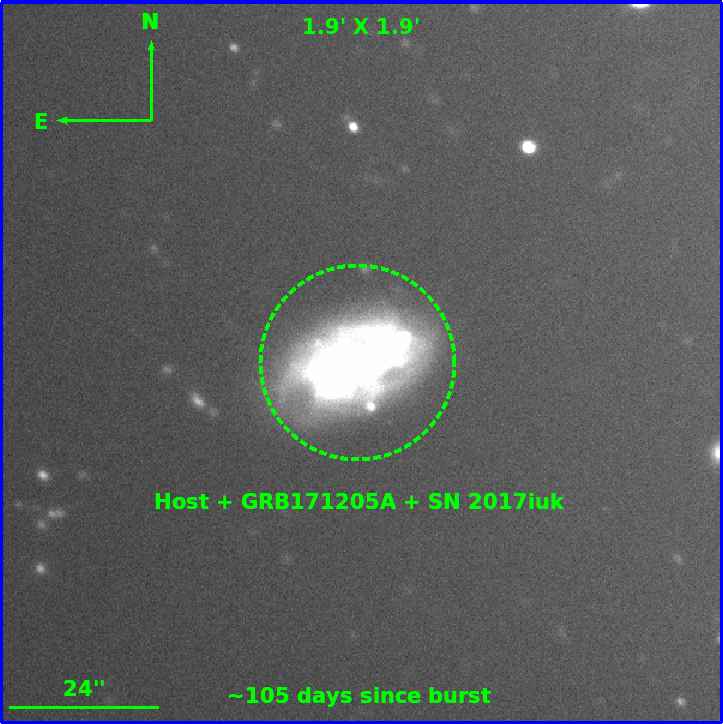}
\includegraphics[angle=0,scale=0.242]{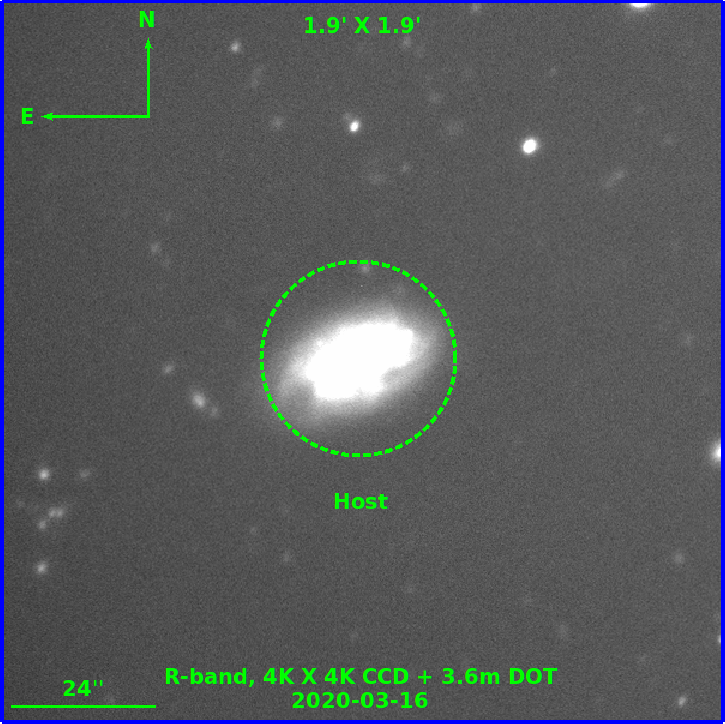}
\includegraphics[angle=0,scale=0.242]{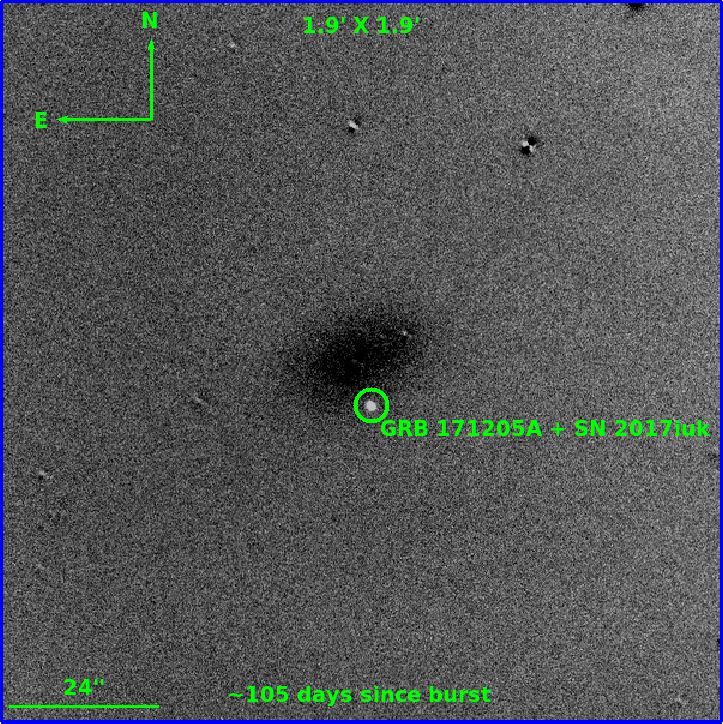}
\caption{The finding chart of GRB~171205A/SN~2017iuk in the $R$-band is presented. The upper-left panel shows a 4.2$^{'}$ $\times$ 4.2$^{'}$ section of the image highlighting GRB171205A+SN 2017iuk+host and seven standard stars (S1--S7) used for calibration. The upper-right panel shows a 1.9$^{'}$ $\times$ 1.9$^{'}$ section of the image, zooming on GRB~171205A/SN~2017iuk associated with the host. The lower-left panel displays the $R$-band image of the field of GRB~171205A/SN~2017iuk observed on 2020-03-16 ($\sim$2.3 yrs post burst), used to perform the template subtraction. The lower-right panel presents the template subtracted image of GRB~171205A/SN~2017iuk, obtained by subtracting the frame of lower-left panel from the image shown in the upper-right panel.}
\label{fig:finding_171205A}
\end{figure*}

\begin{figure*}[ht!]
\centering
\includegraphics[angle=0,scale=0.4]{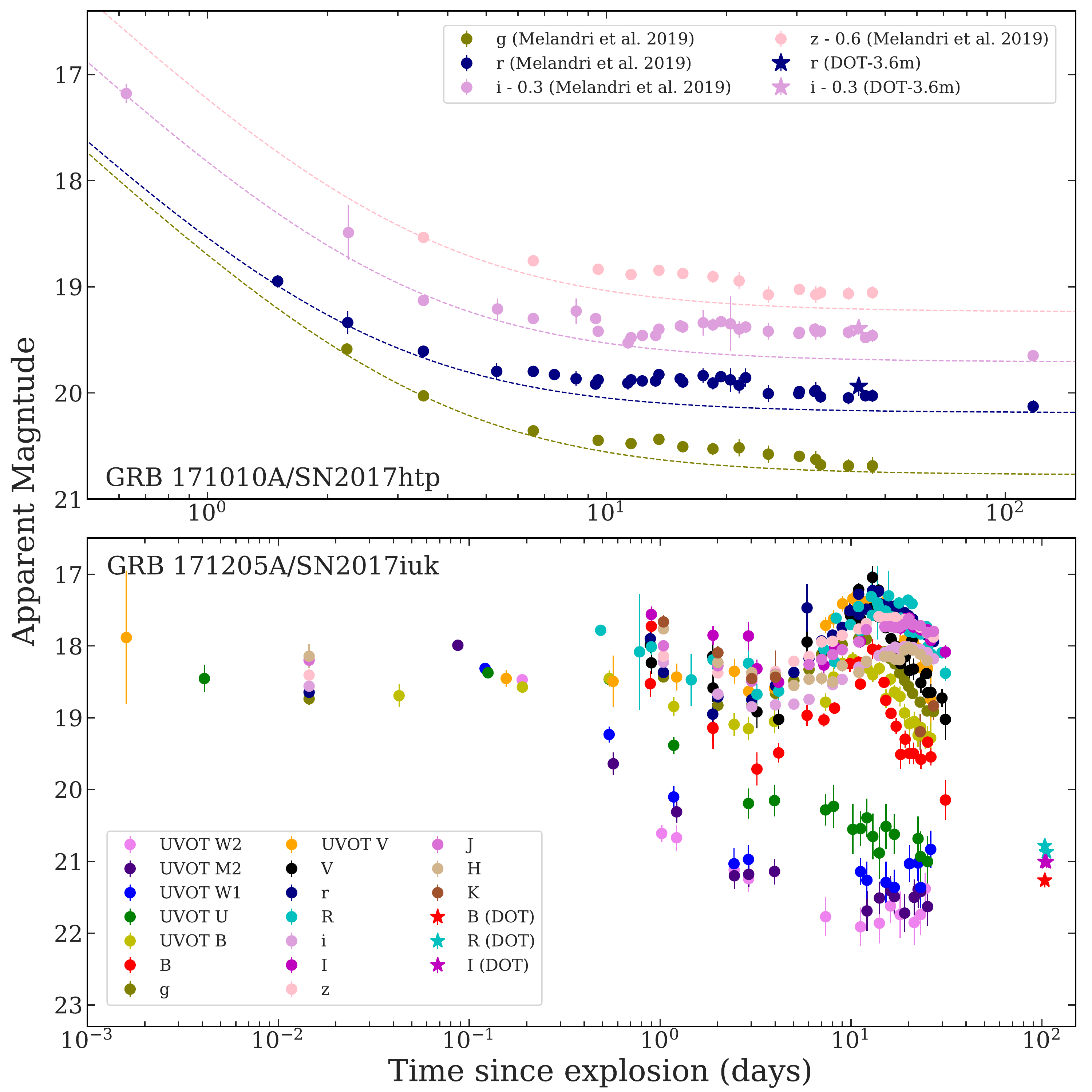}
\caption{Upper panel: multi-band optical light curves of GRB~171010A/SN~2017htp. Data points plotted with star symbols are obtained based on the observations acquired using the 4K$\times$4K CCD Imager and 3.6m DOT, and those plotted with the filled colour-coded circles are taken from \cite{Melandri2019}. The dotted lines present the contribution from the power-law decay of GRB~171010A afterglow and a constant flux from the host galaxy. Lower panel: multi-band light curves (UV--NIR) of GRB~171205A/SN~2017iuk plotted based on the data observed with the 4K$\times$4K CCD Imager/3.6m DOT (with colour-coded star symbols) along with those adopted from \cite{Izzo2019} (with circle symbols).}
\label{fig:LC_magnitudes}
\end{figure*}

\subsection{Optical observations and analysis}
The late-time optical observations of the two events, i.e., GRB~171010A/SN~2017htp and GRB~171205A/SN~2017iuk were obtained using the 4K$\times$4K CCD Imager mounted at the 3.6m DOT \citep{Pandey2018, Kumar2022}, finding charts are shown in Figures~\ref{fig:finding_171010A} and \ref{fig:finding_171205A}, respectively. Observations of GRB/SNe cases presented in this study were acquired with 2$\times$2 binning, a gain of 5 e$^-$/ADU, and a readout speed of 1 MHz, having readout noise of $\sim$10 e$^-$.

\subsubsection{GRB~171010A/SN~2017htp}
The field of GRB~171010A/SN~2017htp was observed in the Bessel $R$-band on 2017-11-21 ($\sim$42d post burst) and detected the underlying GRB-SN embedded in a host galaxy. On 2017-11-22, we again observed the field of GRB~171010A/SN~2017htp in the Bessel $V$, $R$, and $I$ bands. The $R$-band image obtained on 2017-11-22 is shown in the left panel of Figure~\ref{fig:finding_171010A} with a field of view (FOV) of 4.75$'\times$4.75$'$. However, to zoom in on the host embedded GRB~171205A/SN~2017iuk, a smaller section of the GRB field with a FOV of 24$''\times$24$''$ is highlighted in the right panel of Figure~\ref{fig:finding_171010A}. For the field calibration of GRB 171010A/SN 2017htp, Landolt photometric standard fields PG 1047 and SA98 \citep{Landolt1992} were observed on 2021-12-11 along with the GRB field in the $UBVRI$ bands. The standard stars in the PG 1047 and SA98 fields have a $V$-band magnitude range of 11.88 to 15.67 mag and a B$-$V colour range of $-$0.29 to +1.41 mag. For consistency, the data reduction was performed using the procedure discussed in \cite{Melandri2019}, and calibration was done using a standard procedure discussed in \cite{Kumar2021} and utilising the python-scripts hosted on \textsc{RedPipe} \citep{2021redpipe}. The nine secondary standard stars in the GRB field used for the calibration are encircled in the left panel of Figure~\ref{fig:finding_171010A} (S1--S9), whereas their calibrated magnitudes in $UBVRI$ bands are listed in Table~\ref{tab:171010A_secondary_stars}. The complete log of observations of GRB 171010A/SN 2017htp, along with calibrated magnitudes, is tabulated in Table~\ref{tab:Phot_data}. The calibrated $V$, $R$, and $I$-band magnitudes of GRB 171010A/SN 2017htp are converted to the SDSS $r$ and $i$-band magnitudes using the transformation equations of \cite{Jordi2006}. These $r$ and $i$-band magnitudes (with colour-coded star symbols) are plotted with those were reported by \cite{Melandri2019} in $g$, $r$, $i$, and $z$ bands (see the upper panel of Figure~\ref{fig:LC_magnitudes}). The data have also been corrected for the Galactic extinction using E(B-V) = 0.13 mag \citep{Schlafly2011}; however, the host galaxy extinction was considered negligible \citep{Melandri2019}.

\begin{table*}[t]
\caption{Log of multi-band optical observations of GRB~171010A/SN~2017htp and GRB~171205A/SN~2017iuk obtained using the 4K$\times$4K CCD Imager mounted at the axial port of 3.6m DOT. Magnitudes are expressed in the AB system and not corrected for extinction. }\label{tab:Phot_data}
\addtolength{\tabcolsep}{13pt}
\begin{tabular}{c c c c c c}
\hline
JD& $\Delta$t (d)& Filter & Frame  & Mag & Error \\
\hline
\hline
\textbf{GRB 171010A/SN 2017htp}     &    &   &    &    \\
2458079.2916   &  41.999  & R  & 2$\times$100s & 20.24 & 0.06\\ 
               &           &    & 3$\times$300s &  \\ 
2458080.2342   &  42.942  & I  & 6$\times$300s & 19.43 & 0.10\\
2458080.1976   &  42.905  & R  & 8$\times$300s & 20.26 & 0.06\\
2458080.2570   &  42.965  & V  & 6$\times$300s & 20.79 & 0.09\\
\hline
 \textbf{GRB 171205A/SN 2017iuk}    &    &   &     &    \\
2458196.2464  &  103.440 & I  & 3$\times$200s &   21.13 & 0.06 \\ 
2458196.2307  &  103.425 & R  & 2$\times$300s &   20.96 & 0.05 \\
2458196.2384  &  103.432 & B  & 3$\times$200s &   21.56 & 0.10 \\
2458198.2770  &  105.471 & I  & 3$\times$200s &   21.14 & 0.08 \\ 
2458198.2680  &  105.462 & R  & 3$\times$200s &   20.05 & 0.03 \\
\hline
\end{tabular}
\end{table*}

For a typical GRB-SN event, three main flux components contribute at late epochs, i.e., 1) afterglow (AG) of the GRB, 2) the underlying SN, and 3) the constant flux from the host galaxy (see \citealt{Cano2017} for a review). A great deal of physical information about such rare events can be obtained by modelling each of the three components individually. The host galaxy contribution can be extracted through the template-subtraction technique or by subtracting a constant flux of the host using very late time observations. The afterglow component can be separated by fitting either a single or a set of broken power laws to the early time observations. After removing both the host and afterglow contributions, the light curve of the underlying SN can be fetched. In the case of GRB 171010A/SN 2017htp, initially, light curves in all the optical bands followed a single-power law flux decay with $\alpha$ $\approx$ 1.42 $\pm$ 0.05 \citep{Melandri2019}; however, optical light curves started flattening from $\sim$3d post burst. Therefore to remove the contribution of the host galaxy and the afterglow of GRB~171010A, all the presented light curves in the upper panel of Figure~\ref{fig:LC_magnitudes} are fitted with a single power-law having $\alpha$ = 1.42 $\pm$ 0.05 plus a constant to account for the host galaxy contribution (shown with dashed lines) and to extract the light curve of the underlying SN~2017htp.\\

\subsubsection{GRB~171205A/SN~2017iuk}
The field of GRB~171205/SN~2017iuk was observed on 2018-03-18 and 2018-03-20 ($\sim$103 and 105d post burst) respectively in $BRI$ and $RI$ bands using the 4K$\times$4K CCD Imager at the 3.6m DOT. We again observed the field on 2020-03-16 (in $RI$ bands; $\sim$2.3 yrs post burst to perform image subtractions) and 2021-01-14 (in $UBVRI$ bands, $\sim$3.1 yrs post burst for image subtraction and calibrations). For the purpose of field calibrations, Landolt photometric standard field PG~1323 \citep{Landolt1992} was also observed on 2021-01-14 along with the GRB field in the $UBVRI$ bands. The standard stars in PG~1323 have a $V$-band magnitude range of 12.09 to 14.00 mag and a B$-$V colour range of $-$0.14 to +0.76 mag. The $R$-band finding chart of GRB~171205A/SN~2017iuk and secondary standard stars (S1--S7) with a FOV of 4.2$'\times$4.2$'$ are shown in the upper-left panel of Figure~\ref{fig:finding_171205A}. A smaller section with a FOV of 1.9$'\times$1.9$'$ of GRB 171205A/SN 2017iuk field is presented in the upper-right panel. The frame used for the template subtraction and the template subtracted image are shown respectively in the lower left and right panels of Figure~\ref{fig:finding_171205A}. The template subtraction to remove the host galaxy contributions was performed with a standard procedure via matching FWHM and flux values of respective images using the python-based codes hosted on \textsc{RedPipe} \citep{2021redpipe}. Calibrated $UBVRI$ magnitudes of seven secondary standard stars of the GRB/SN field are tabulated in Table~\ref{tab:secondary_stars_aa}, whereas calibrated magnitudes of the GRB~171205A/SN~2017iuk thus obtained are tabulated in Table~\ref{tab:Phot_data}. These calibrated magnitudes are plotted (with star symbols) along with publicly available ones in $W2M2W1UBgVrRiIzJHK$ bands (with colour-coded circle legends) as reported by \citealt{Izzo2019} (see the lower panel of Figure~\ref{fig:LC_magnitudes}). The data have also been corrected for the host extinction of E(B-V) = 0.02 mag \citep{Izzo2019} and a Galactic extinction of E(B-V) = 0.05 mag \citep{Schlafly2011}.
 
Since the beginning, optical afterglows of GRB~171205A have deviated from the power-law decay; however, light curves are completely dominated by the BB emission from SN~2017iuk after $\gtrsim$3d post-burst. Hence we considered the GRB~171205A afterglow contribution negligible at late times. The SN optical light-curve peak is observed around $\sim$11d post burst \citep{Postigo2017, Izzo2019}. In the case of GRB~171205A/SN~2017iuk, the host galaxy component is removed via performing template subtraction.

\section{Results}\label{sec:results}
This section describes major results emphasising the importance of prompt emission properties and underlying SNe properties of the two GRB-SNe in the optical and other similar objects. 

\subsection{High energy aspects}

\subsubsection{Prompt emission spectra}
\label{promptspectra}

\begin{figure*}[!t]
\centering
\includegraphics[scale=0.4]{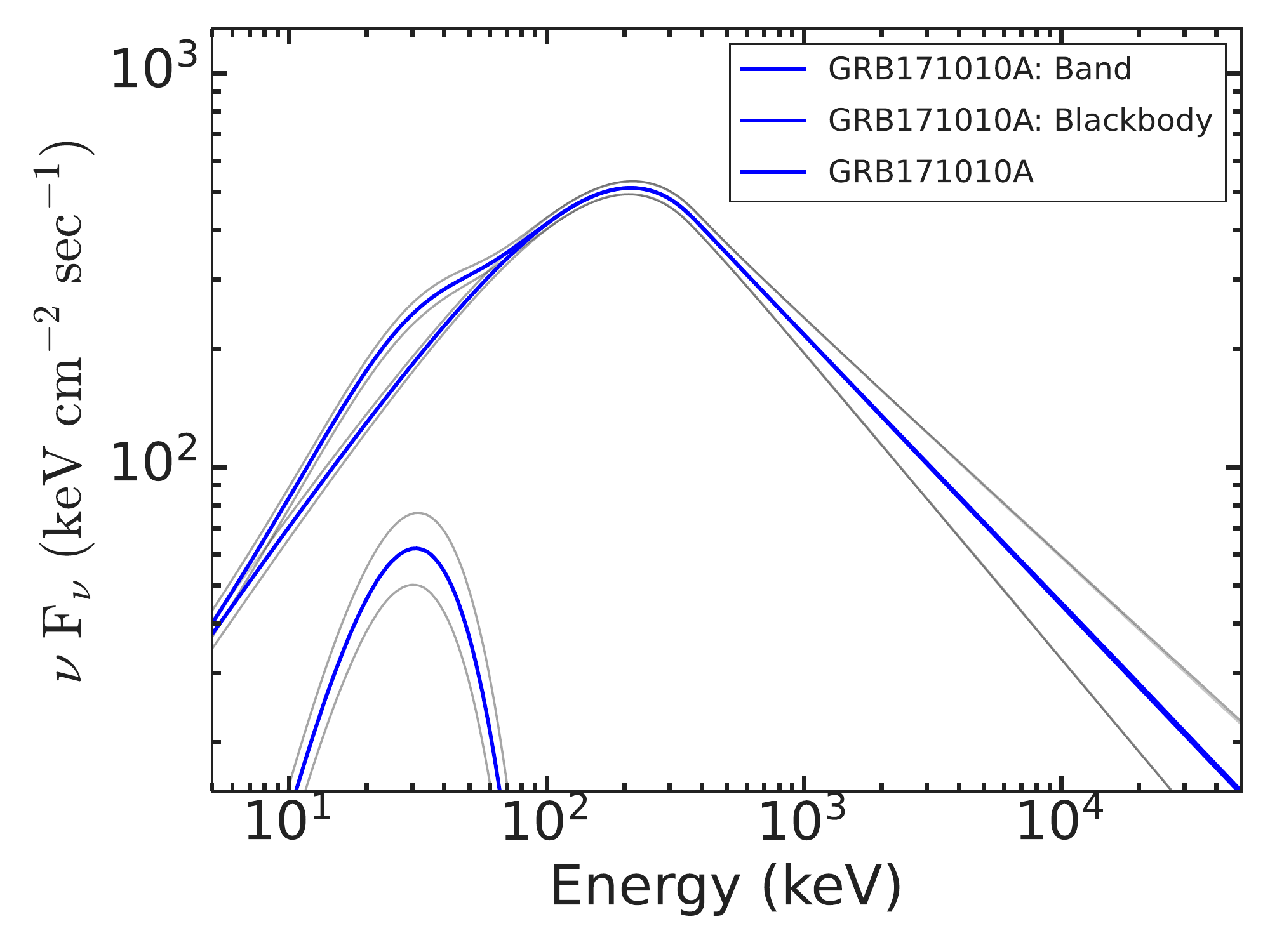}
\includegraphics[scale=0.17]{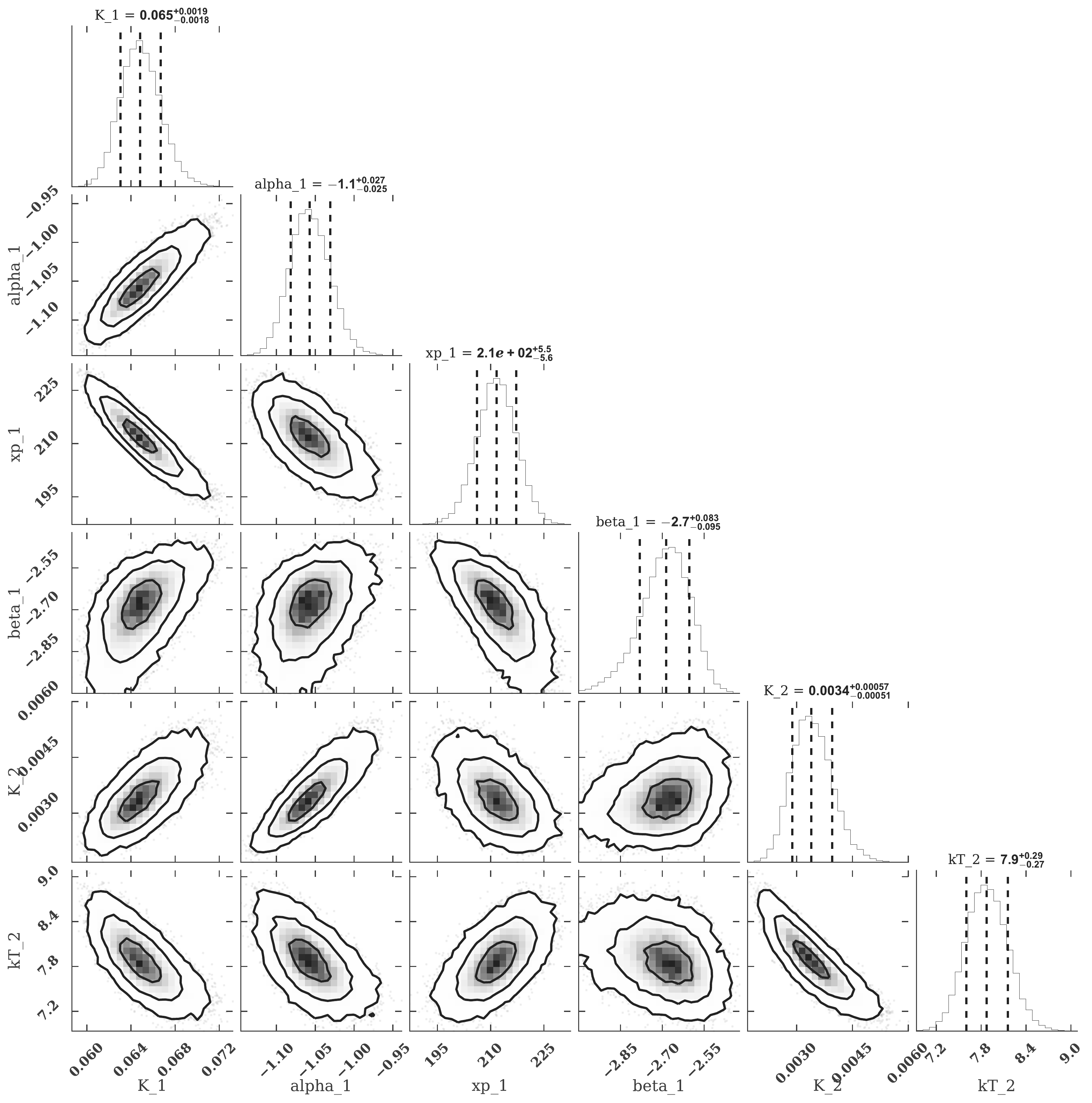}
\includegraphics[scale=0.4]{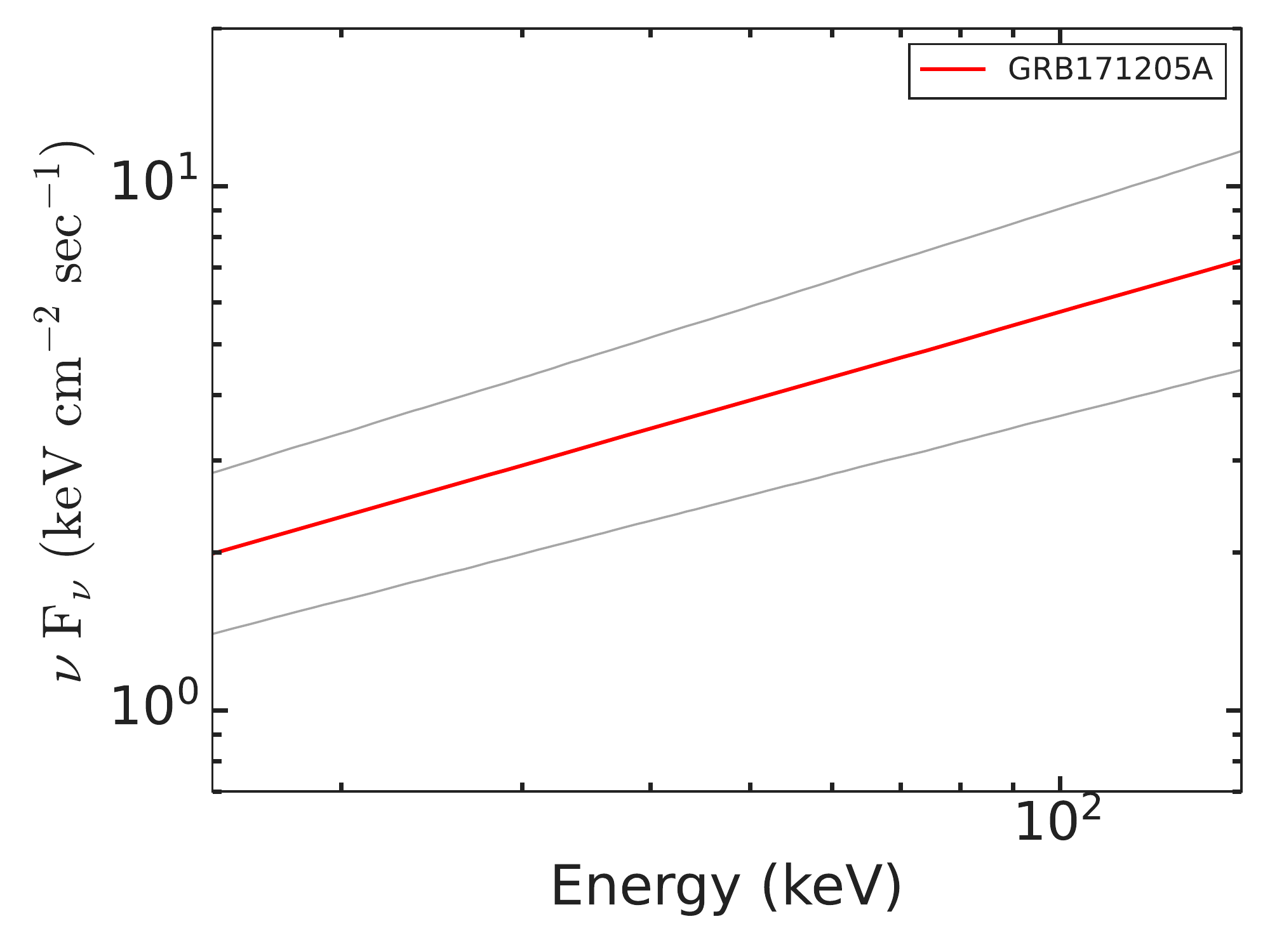}
\includegraphics[scale=0.5]{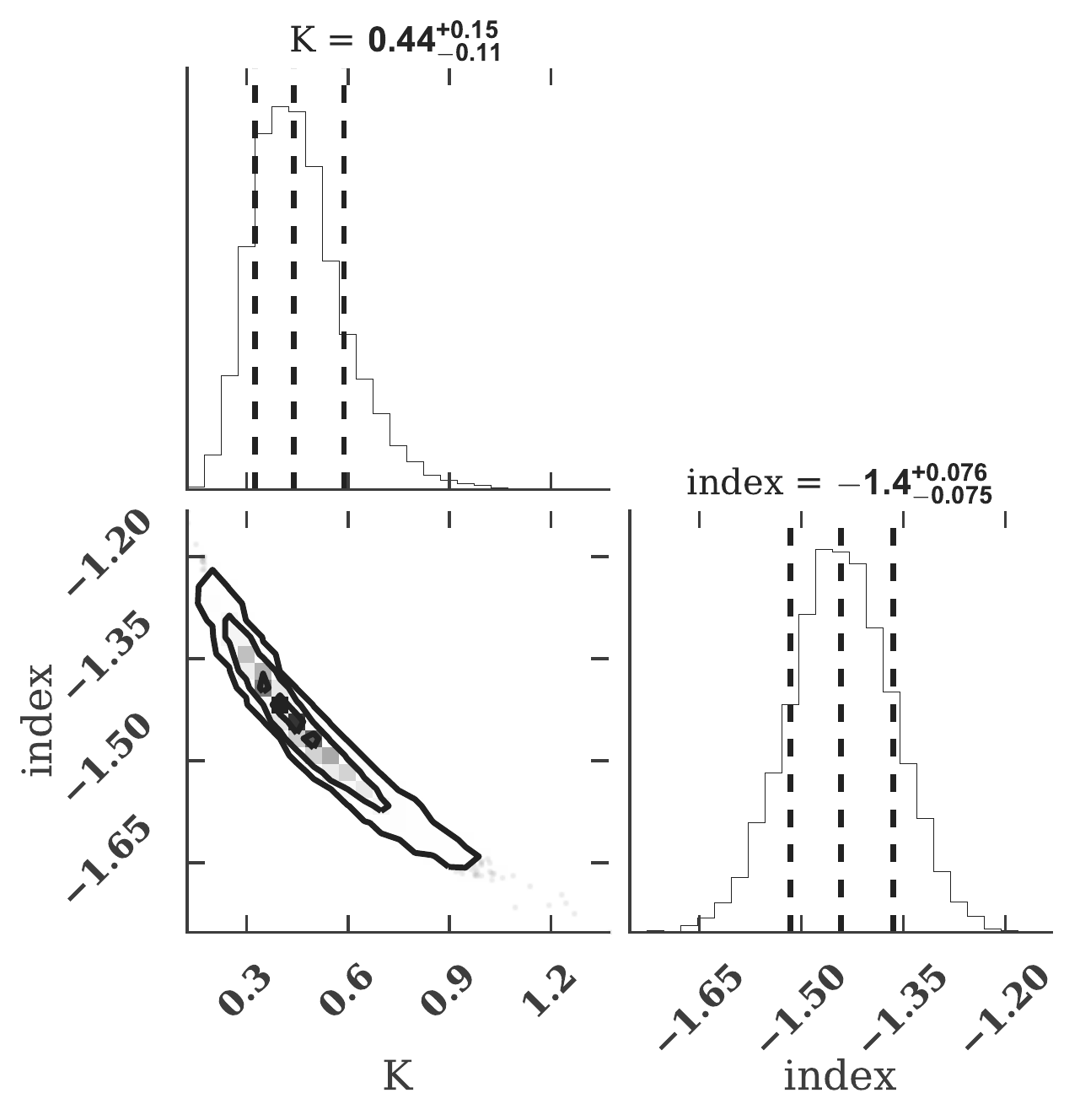}
\caption{{\it Time-averaged spectra$-$} Upper panel: the best fit \fermi GBM time-averaged (from \fermiT-5 to \fermiT +278s) spectrum of GRB~171010A. Lower panel: the best fit \swift BAT time-averaged (from \swiftT -42.228 to \swiftT +197.796s) spectrum of GRB~171205A. The corner plots obtained using 2000 number of simulations using Bayesian fitting are also shown for both bursts.}\label{batspectrum}
\end{figure*}

\begin{figure}[!t]
\centering
\includegraphics[scale=0.34]{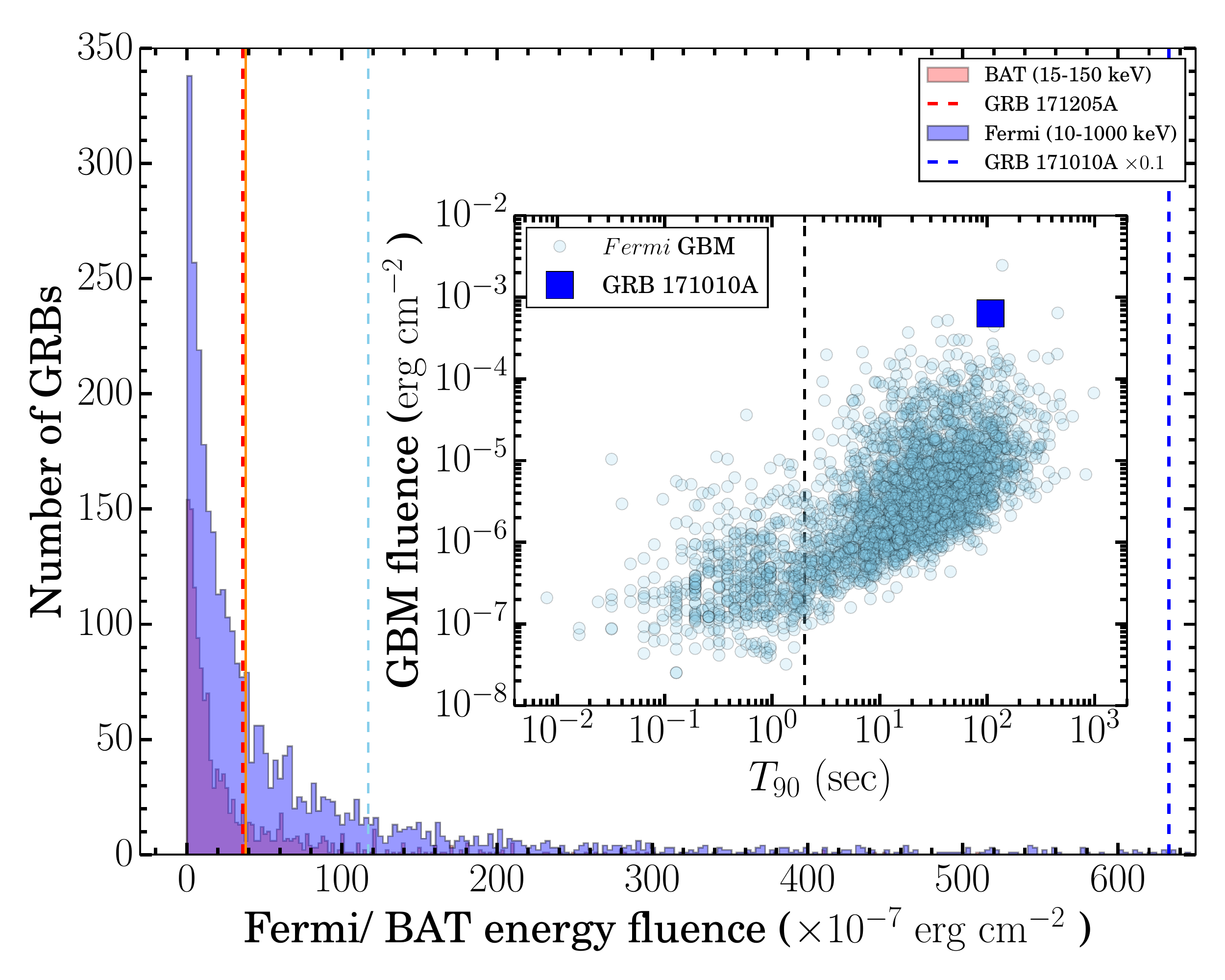}
\caption{{\it Observed fluence distribution$-$} The fluence distributions for GBM (blue) and BAT (red) detected GRBs in the 10-1000 and 15-150 \keV energy ranges, respectively. The vertical solid dark orange and dashed sky blue lines show the mean values of the BAT and GBM samples, respectively. The positions of GRB~171010A and GRB~171205A are shown with blue and red dashed vertical lines, respectively. The inset diagram presents the energy fluence as a function of \tninty distribution of GBM detected bursts. The position of GRB~171010A is shown with a blue square. The vertical black dashed line corresponds to the \tninty = 2s.}\label{bat_ppf_fluence}
\end{figure}

The \fermi GBM time-integrated spectrum of GRB~171010A (in model space, from \fermiT-5 to \fermiT+278 s), along with the corner plot, is shown in the upper panel of Figure~\ref{batspectrum}. It can be well explained using the \sw{Band} plus \sw{Blackbody} model.  We obtained following spectral parameters for the best fit model: low-energy photon index ($\alpha$) = -1.06 $\pm$ 0.03, high-energy photon index ($\beta$) = 2.69 $\pm$ 0.09, peak energy (\Ep) = 211.69 $\pm$ 5.55 \keV and temperature kT = 7.88 $\pm$ 0.28 \keV. The $\alpha$ value is consistent with the prediction of synchrotron emission in slow and fast cooling cases. However, the detection of a thermal (photospheric) signature in combination with a non-thermal \sw{Band} suggests a hybrid jet composition (a matter-dominated hot fireball and a colder magnetic-dominated Poynting flux component) for GRB~171010A. We compared the observed energy fluence of GRB~171010A with all GBM detected GRBs (see Figure \ref{bat_ppf_fluence}) and proposed that GRB~171010A is the third most GBM fluent burst.

We also performed the \swift BAT time-averaged (from \swiftT -42.228 to \swiftT +197.796 s) spectral analysis of GRB~171205A using the \sw{3ML} tool and Bayesian fitting. We noticed that the \swift BAT time-averaged spectrum of GRB~171205A is best fitted by the power-law model (lowest BIC value) with a photon index of 1.44 $\pm$ 0.08. The best fit spectrum, along with the corner plot, is shown in the lower panel of Figure \ref{batspectrum}. During the time-integrated temporal window, we estimated the fluence equal to $\approx$(3.6 $\pm$ 1.6) $\rm \times 10^{-6} erg~cm^{-2}$ in 15-150 \keV energy range. Once compared with the distribution of the BAT fluence in 15-150 \keV for all the \swift BAT detected bursts, we find that the energy fluence of GRB~171205A is close to the mean value $\approx$(3.8 $\rm \times 10^{-6} erg~cm^{-2}$) of the sample (see Figure \ref{bat_ppf_fluence}).

\subsubsection{Prompt emission characteristics}

\begin{figure}[!t]
\centering
\includegraphics[scale=0.31]{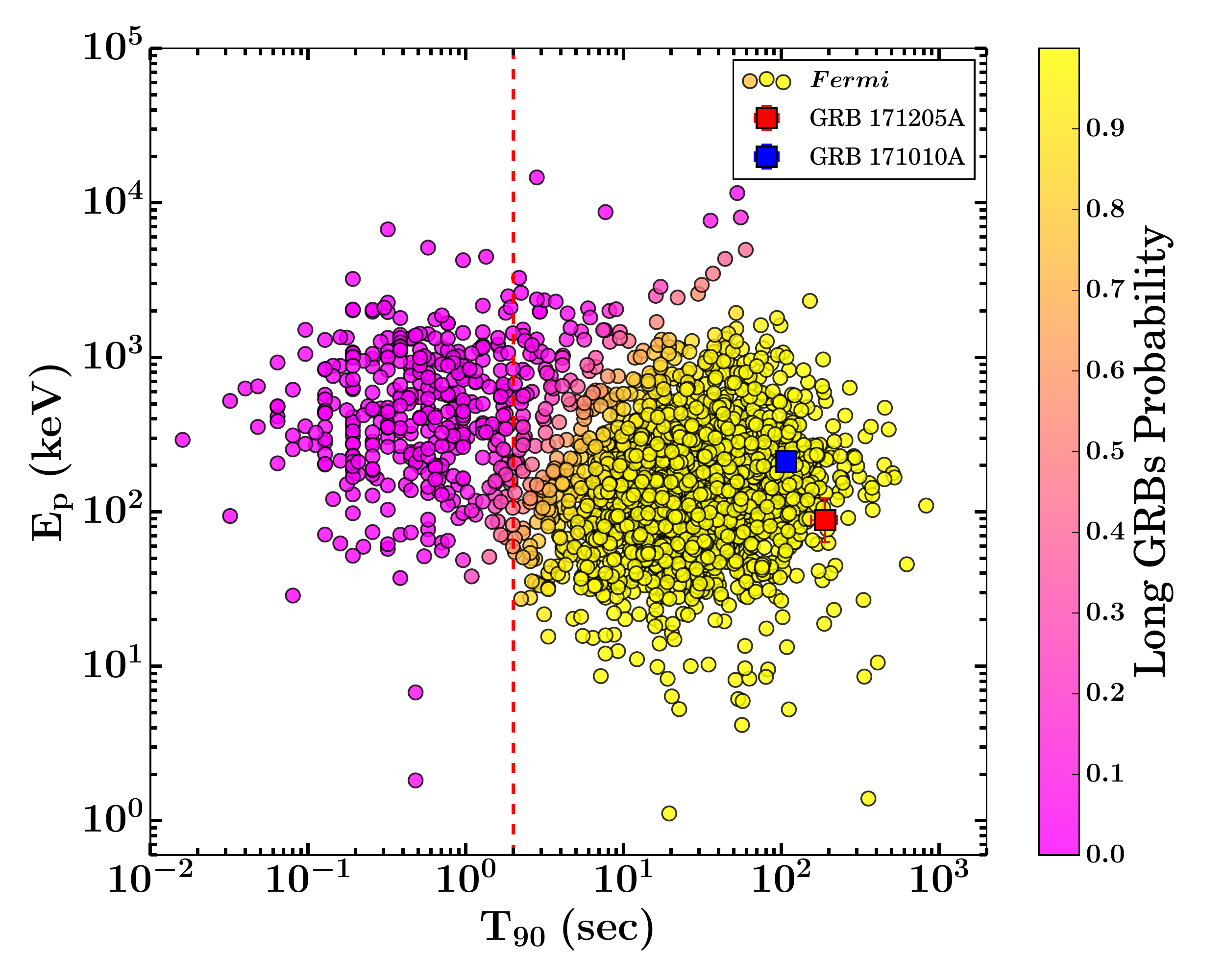}
\includegraphics[scale=0.31]{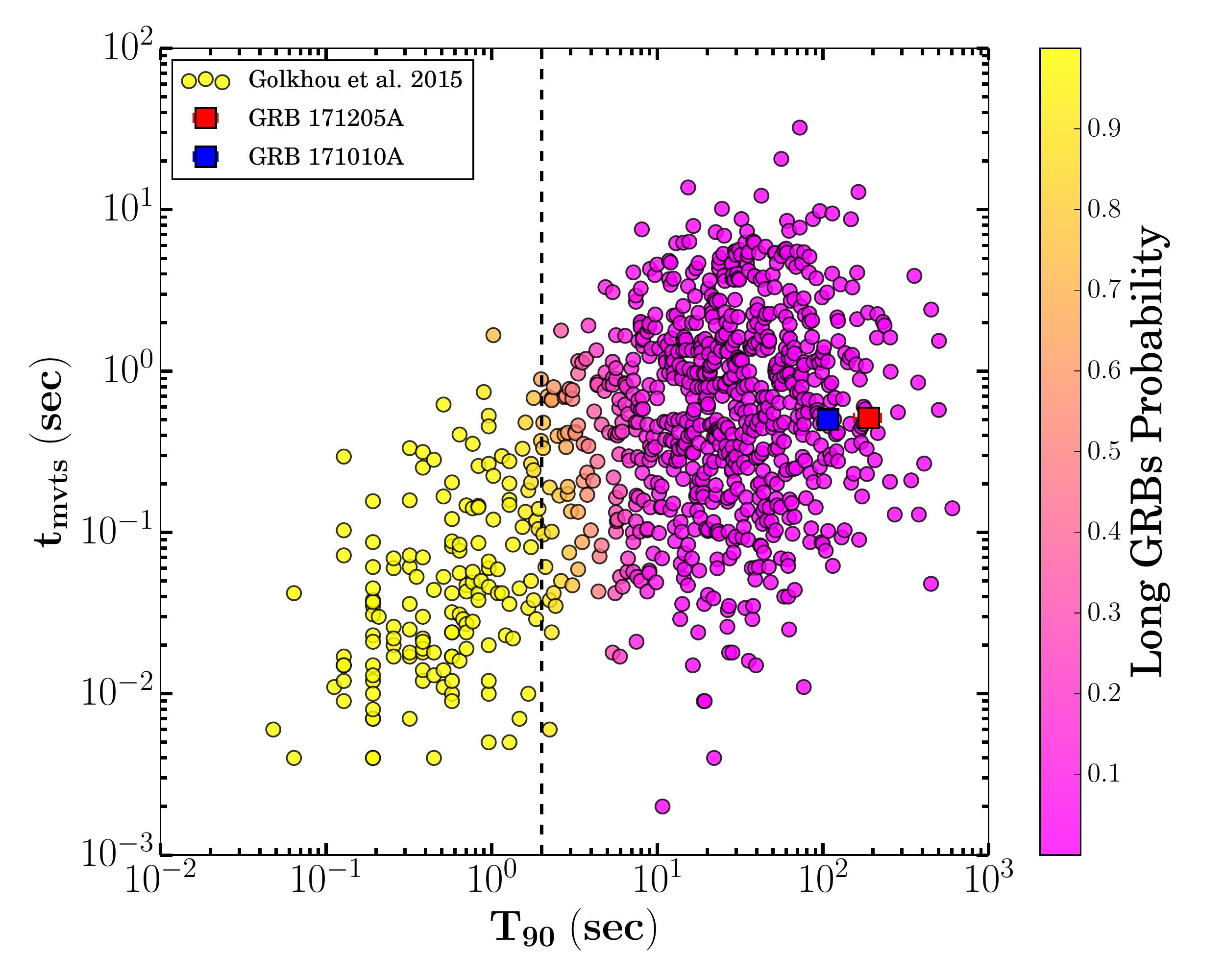}
\caption{{\it Spectral hardness$-$} Left panel: \Ep-\tninty diagram. A sample of long and short GRBs taken from the \fermi GBM catalogue is shown. The positions of GRB 171010A and GRB 171205A are highlighted with blue and red squares, respectively. The right side Y-scale represents the probability of lGRBs. The vertical red dashed line shows the line of the traditional classification of GRBs. Right panel: \mvts-\tninty diagram. The \mvts versus \tninty for GRB~171010A (blue square) and GRB~171205A (red square) are plotted with a sample of long and short GRBs adopted from \protect\cite{2015ApJ...811...93G} are presented. The vertical black dashed line guides the eyes to look at the traditional classification of GRBs.}\label{t90HR}
\end{figure}

\paragraph{{\it Spectral peak-duration distribution:}}
The lGRBs are believed to have a softer spectrum in comparison to those observed for the majority of sGRBs \citep{1993ApJ...413L.101K}. We determined the time-integrated spectral \Ep (hardness) of GRB~171010A and GRB~171205A and placed them in \Ep-\tninty plane to compare with a larger sample of bursts detected by the \fermi GBM (see the left panel of Figure \ref{t90HR}). We found that both GRBs have softer \Ep values and durations longer than 100s. GRB~171010A and GRB~171205A lie towards the right edge in the spectral peak-duration distribution plane and follow the typical characteristics of lGRBs.\\

\paragraph{{\it Minimum Variability Time scale:}} We determined the minimum variability time scales (\mvts) of GRB~171010A and GRB~171205A using the Bayesian block algorithm on count rate prompt emission light curve in GBM 8-900 \keV and BAT 15-350 \keV, respectively (see Figure \ref{batlc}). The Bayesian blocks follow the statistically significant changes in the count rate light curves of GRB~171010A and GRB~171205A. We calculated the minimum bin size of the Bayesian blocks. The half of the minimum bin size of Bayesian blocks is considered as \mvts \citep{2018ApJ...864..163V}. The right panel of Figure \ref{t90HR} shows the \mvts-\tninty distribution for a sample of long and short GRBs studied by \cite{2015ApJ...811...93G}. We found that the \mvts values of both the bursts are nearly comparable. The positions of GRB~171010A and GRB~171205A are shown with blue and red squares, respectively, and they lie towards the right edge of the distribution. We also calculated the probability of lGRBs of the sample studied by \cite{2015ApJ...811...93G} using the Gaussian mixture model.  

\subsubsection{Prompt Correlation$-$Amati}

We examine the correlation between the rest-frame \Ep and isotropic equivalent gamma-ray energy ($E_{\rm \gamma, iso}$) for GRB~171010A and GRB~171205A in the Amati plane \citep{2006MNRAS.372..233A}. In the case of GRB~171010A, we calculated the \Ep and $E_{\rm \gamma, iso}$ of the burst from the time-integrated spectral modelling of GBM data. As the \swift BAT instrument has a limited spectral coverage (15-150 \keV), therefore, in the case of GRB~171205A, we calculated the time-integrated \Ep using the correlation between the observed BAT fluence and \Ep \citep{2020ApJ...902...40Z}, i.e., \Ep = [fluence/($\rm 10^{-5} erg~cm^{-2}$)]$^{0.28}$ $\rm \times 117.5^{+44.7}_{-32.4}$ \keV $\approx$ 88.27$^{+33.58}_{-24.34}$ \keV, a softer value of \Ep, typically observed for lGRBs. The calculated \Ep value of GRB~171205A is consistent with that reported by \cite{DElia2018} using joint \kw and BAT observations. To estimate the $E_{\rm \gamma, iso}$ for GRB~171205A, we used equation 6 of \cite{2015ApJ...815..102F}. 

\begin{figure}[!t]
\centering
\includegraphics[scale=0.35]{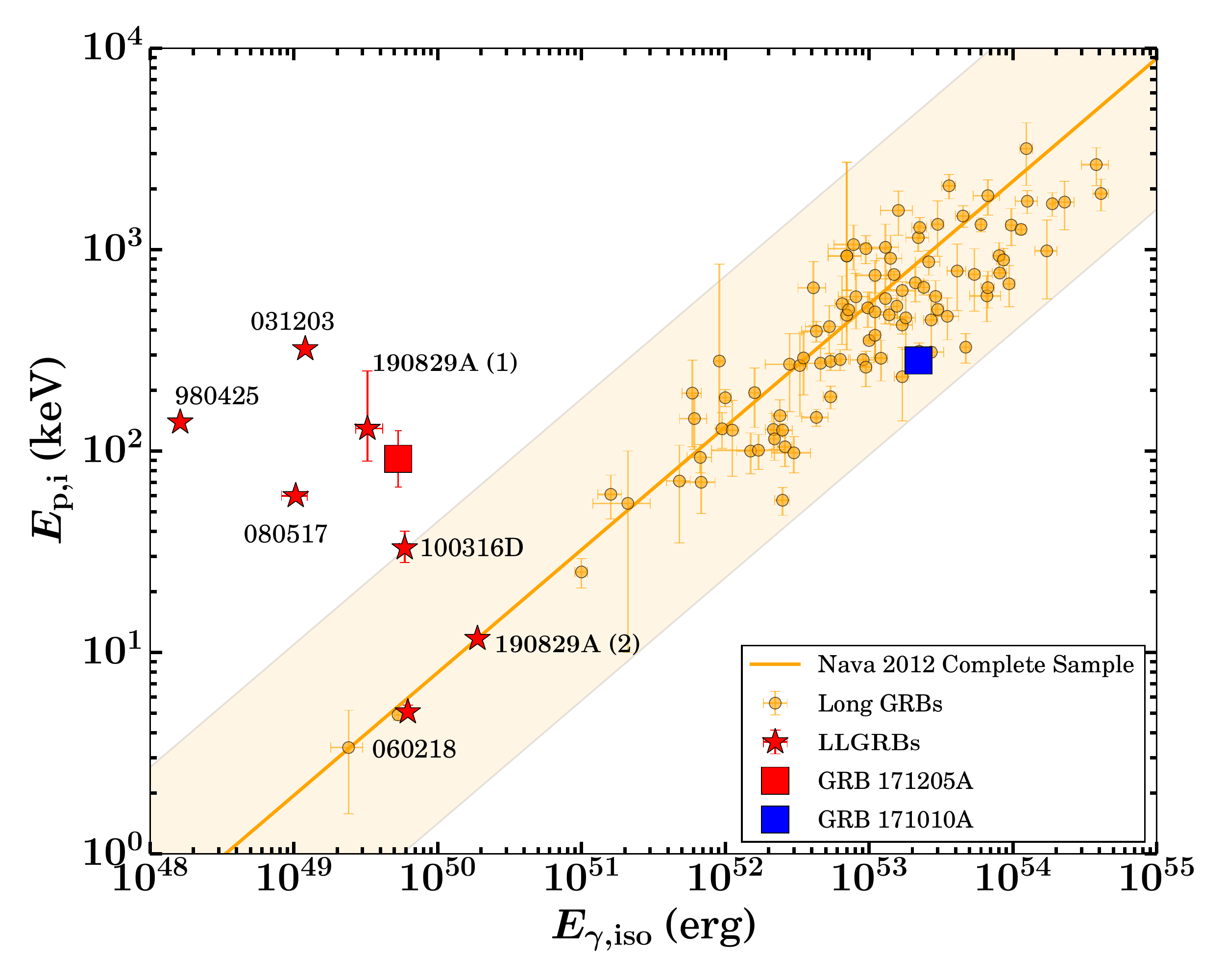}
\caption{{\it Amati correlation:} GRB~171010A (blue square) and GRB~171205A (red square) in Amati correlation along with the samples of lGRBs \protect\cite[taken from;][]{2020MNRAS.492.1919M} and llGRBs \protect\citep[adopted from;][]{2020ApJ...898...42C}. The figure clearly indicates that GRB~171010A is well consistent with the Amati correlation, whereas GRB~171205A is an outlier to this correlation, as seen in the case of other well-studied llGRBs.}
\label{amati}
\end{figure}

The calculated $E_{\rm \gamma, iso}$ values suggest that GRB~171010A is a luminous burst \citep{Chand2019}; however, GRB~171205A is a llGRB. Amati correlation for GRB~171010A and GRB~171205A along with a sample of lGRBs taken from \cite{2020MNRAS.492.1919M} and a set of llGRBs adopted from \cite{2020ApJ...898...42C} is shown in Figure \ref{amati}. We noticed that despite being connected to SNe, GRB~171205A is an outlier to the Amati correlation \citep{DElia2018}, similar to other llGRBs; on the other hand, GRB~171010A is well consistent with this correlation. Generally, llGRBs are not agreeable with the Amati correlation as the radiation from these sources is powered by shock breakout \citep{2012ApJ...747...88N, 2015MNRAS.448..417B, 2020ApJ...898...42C} among other alternate reasons. For llGRBs, the energetics satisfy a fundamental correlation which is helpful in constraining the observed duration of the burst \citep{2012ApJ...747...88N}. We have used equation 18 of \cite{2012ApJ...747...88N} to estimate the observational duration for GRB~171205A due to shock breakout and found $\sim$ 300 s, which is in agreement with the observed \tninty duration of the burst within errors. This indicates that GRB~171205A might be originated due to shock breakout, as seen for a few other llGRBs \citep{2012ApJ...747...88N, 2020ApJ...898...42C}. Alternatively, \cite{DElia2018} discussed that GRB~171205A lies outside the Amati correlation based on joint \swift and \kw observations and using a different spectral analysis method. They also mentioned that GRBs observed off-axis could not be consistent with the Amati correlation, whereas GRBs with on-axis observations generally follow this correlation. In the case of GRB~171205A, the source was observed off-axis; it might also be a possible cause for the burst to be an outlier in the Amati correlation. Further, \cite{2017A&A...608A..52M} suggested that if an old generation instrument observes a particular GRB (i.e., GRB 980425 and GRB 031203), that might cause a false outlier of Amati correlation, although this possibility is less likely for GRB~171205A as \swift (new generation instrument) and \kw both confirmed that the burst is an Amati outlier. In the light of the above, we suggest that GRB 171205A could not be an outlier to the Amati correlation due to the observational biases rather than the intrinsic nature of typical llGRBs (shock breakout or off-axis view).

\subsubsection{Possible central engines}\label{sec:central_engine}

A central engine powering source is widely accepted to explain the formation of lGRBs. It could be in the form of a mass accreting black hole \citep{Woosley1993, MacFadyen1999} or a millisecond magnetar \citep{Duncan1992, Usov1992, Thompson1993}. To constrain the possible central engines of GRB~171010A and GRB~171205A, we used methods presented in \cite{2021ApJ...908L...2S}. The key idea of this method is based on the maximum possible rotational energy of the order of $\sim$10$^{52}$ erg to launch jets from a millisecond magnetar. For this purpose, we determined the jet opening angles and beaming-corrected gamma-ray energy ($E_{\rm \gamma, beamed}$) for both the bursts. We constrained lower limits on the jet opening angle using equation 4 given in \cite{2001ApJ...562L..55F}. We assumed typical values of number density of medium ($n_{0}$ = 1) and thermal energy fraction in electrons ($\epsilon_{e}$ = 0.1) to calculate the jet opening angles \citep{2002ApJ...571..779P, 2021arXiv211111795G}. 

\begin{figure}[!t]
\centering
\includegraphics[scale=0.32]{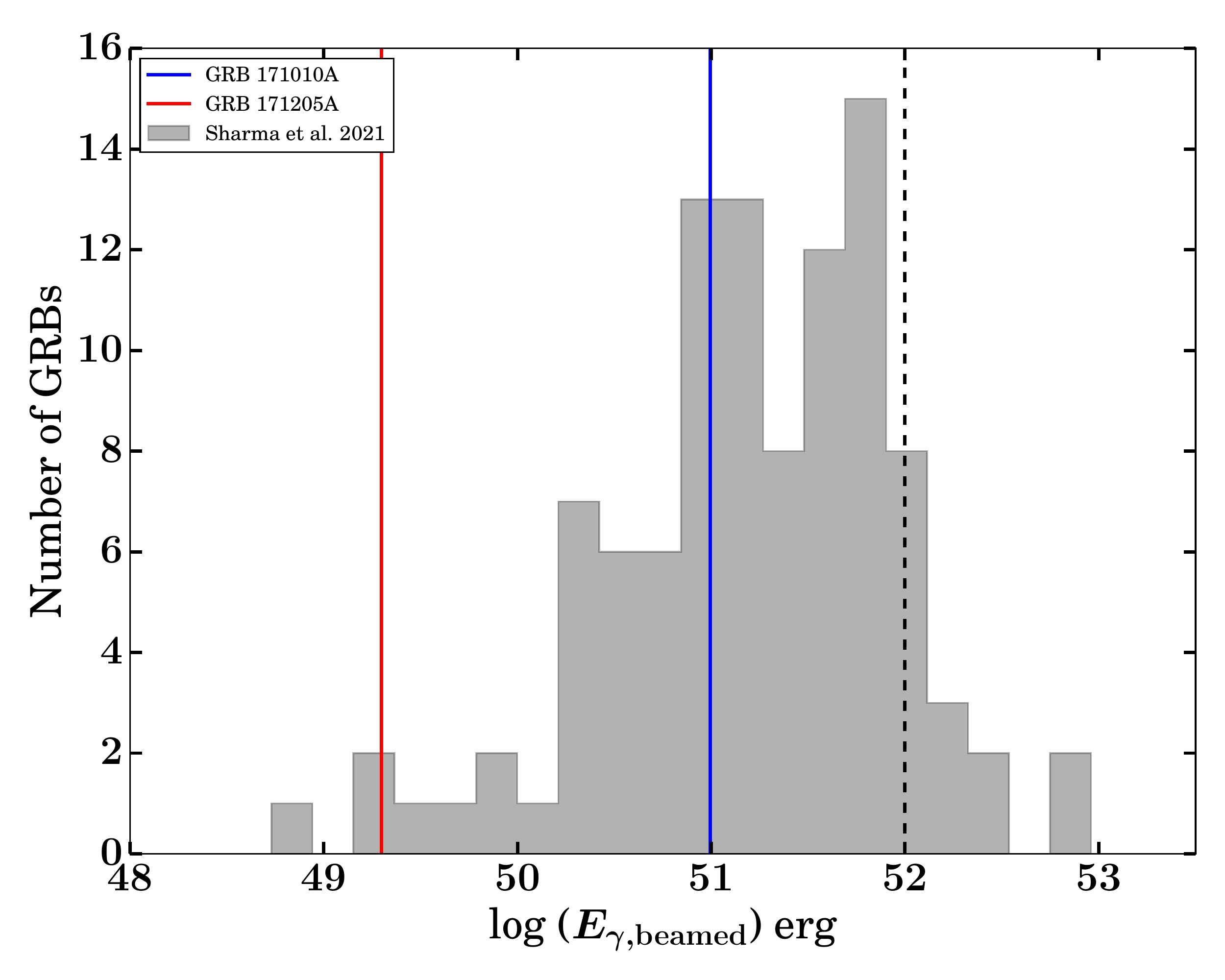}
\includegraphics[scale=0.32]{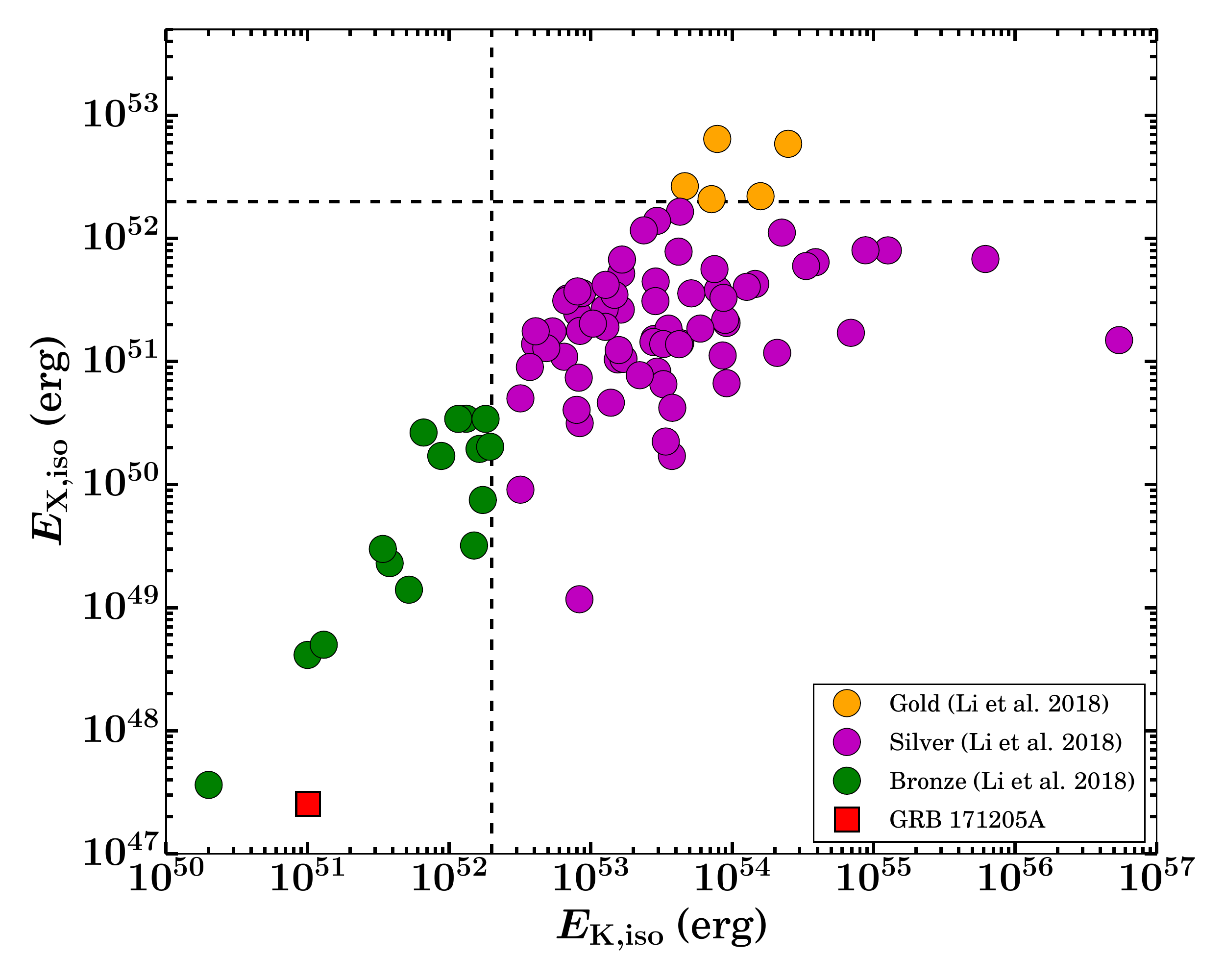}
\caption{{\it Possible central engine$-$} Left panel: distribution of $E_{\rm \gamma, beamed}$ for all GRBs detected by GBM, taken from \protect\cite{2021ApJ...908L...2S}. The blue and red lines highlight the respective positions of GRB~171010A and GRB~171205A. The vertical black line shows the maximum possible rotational energy of the magnetar. Right panel: the distribution of $E_{\rm X, iso}$ and $E_{\rm K, iso}$ for gold (orange), silver (magenta), and bronze (green) samples of \protect\cite{Liang2018} is shown. The position of GRB~171205A is marked with a red square. The vertical and horizontal black dashed lines show the maximum possible energy budget of a magnetar as a central engine powering source.}
\label{central_engine} 
\end{figure}

In the case of GRB~171010A, we considered second temporal break as jet break time (see Section \ref{XRTafterglow}) and found $\theta_{j}$ = 5.43$^\circ$ translated to $E_{\rm \gamma, beamed}$  equal to 9.91 $\times ~10^{50}$ erg. On the other hand, in the case of GRB~171205A, we used the last XRT data points to constrain the limits on the jet opening angle. We found a very wide opening angle $\theta_{j}$ $>$ 51.3$^\circ$ and $E_{\rm \gamma, beamed}$ $>$ 1.99 $\times ~10^{49}$ erg. We noticed that for both the cases, $E_{\rm \gamma, beamed}$ values are less than the maximum possible rotational energy budget of a typical magnetar \citep{2009A&A...502..605H}. 
Therefore, this method indicates that the magnetars could be the central engine for the two bursts under discussion. The existence of a Kerr Black Hole as an inner engine demands a higher energy budget among other observed properties \citep{2017MNRAS.464.3219V, 2021ApJ...908L...2S}. The left panel of Figure \ref{central_engine} shows the number distribution of $E_{\rm \gamma, beamed}$ of all the GBM detected bursts taken from \cite{2021ApJ...908L...2S}. The locations of GRB~171010A and GRB~171205A are also marked using blue and red lines, respectively.

We further utilised the method discussed by \cite{Liang2018} to constrain the possible central engine of GRB~171205A investigating an $X$-ray plateau followed by a normal decay phase. \cite{Liang2018} performed the analysis of the $X$-ray afterglow light curve having plateau phases of 101 bursts (up to May 2017). They measured redshift values by calculating the isotropic $X$-ray ($E_{\rm X, iso}$) and kinetic ($E_{\rm K, iso}$) energies and compared them with the maximum energy budget of magnetars. In the case of GRB~171205A, we calculated the $E_{\rm X, iso}$ (2.59 $\times ~10^{47}$ erg) using equation 7 of \cite{Liang2018}. Further, we followed the methodology discussed by \cite{Liang2018} to calculate the $E_{\rm K, iso}$ and have applied the closure relations during the normal decay phase of GRB~171205A\footnote{\url{https://www.swift.ac.uk/xrt_live_cat/00794972/}} to constrain the spectral regime. We found that the normal decay phase could be explained if the observed frequencies $\rm \nu_{X} > max (\nu_{c},\nu_{m})$; where $\nu_{c},\nu_{m}$ are the cooling and maximum synchrotron break frequencies for the external forward shock model both for ISM and wind like ambient media. We used equation 9 of \cite{Liang2018} to calculate $E_{\rm K, iso}$ (1.01 $\times ~10^{51}$ erg) for GRB~171205A; for micro-physical parameters: $\epsilon_{e}$ = 0.1, thermal energy fraction in magnetic filed ($\epsilon_{B}$) = 0.01, $n_{0}$ = 1, and the Compton parameter (Y) = 1. The distribution of $E_{\rm X, iso}$ and $E_{\rm K, iso}$ for the defined gold, silver, and bronze samples taken from \cite{Liang2018} are shown in the right panel of Figure \ref{central_engine}. GRB~171205A lies towards the lower-left edge of the distribution, and its position is marked with a red square. The measured values of $E_{\rm X, iso}$ and $E_{\rm K, iso}$ for GRB~171205A are less than the maximum energy budget of magnetars and belong to the bronze sample of \cite{Liang2018}, providing a clue that magnetar could be the possible central engine for GRB~171205A. However, we could not implement this method for GRB~171010A due to the non-existence of the plateau phase followed by a normal decay phase in the XRT light curve.

\begin{figure}[!t]
\centering
\includegraphics[scale=0.75]{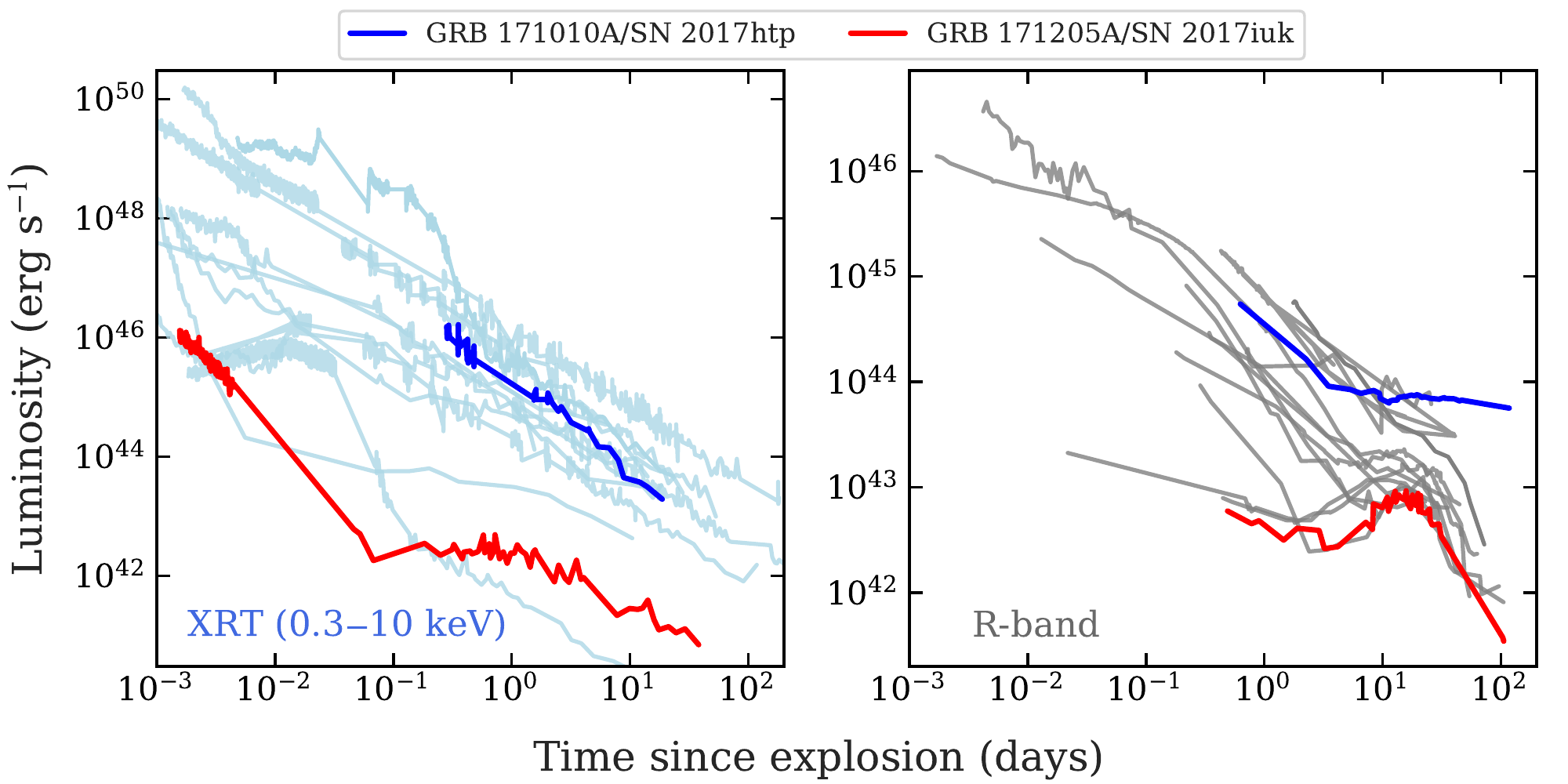}
\caption{{\it $X$-ray and $R$-band afterglow light curves$-$} the XRT and optical $R$-band light curves of GRB~171010A/SN~2017htp (shown with blue) and GRB~171205A/SN~2017iuk (shown with red) along with the XRT-observed light curves of all known GRB/SNe cases are shown in the left and right panels, respectively. GRB~171010A/SN~2017htp  and GRB~171205A/SN~2017iuk respectively appear among the brightest and faintest GRB/SNe.}
\label{XRT}
\end{figure}

\subsection{$X$-ray and optical light curves comparison}
\label{XRTafterglow}
In this section, $X$-ray and optical $R$-band light curves of GRB~171010A/SN~2017htp and GRB~171205A/SN~2017iuk are compared with those of 13 other well studied XRT-detected GRB/SNe cases (see the left and right panels of Figure~\ref{XRT}, respectively). The comparison sample includes : GRB 050525A/SN 2005nc (@$z$ = 0.606), GRB 081007A/SN 2008hw (@$z$ = 0.530), GRB 091127A/SN 2009nz (@$z$ = 0.490), GRB 101219B/SN 2010ma (@$z$ = 0.552), GRB 111209A/SN 2011kl (@$z$ = 0.677), GRB 130702A/SN 2013dx (@$z$ = 0.145), GRB 130831A/SN 2013fu \citep[@$z$ = 0.479;][and references therein]{Cano2017}, GRB 060218/SN 2006aj \citep[@$z$ = 0.033;][]{Ferrero2006}, GRB 120422A/SN 2012bz \citep[@$z$ = 0.282;][]{Schulze2014}, GRB 130427A/SN \citep[@$z$ = 0.340;][]{Perley2014}, GRB 190114C/SN \citep[@$z$ = 0.425;][]{Misra2021, 2022MNRAS.511.1694G}, GRB 190829A/SN 2019oyw \citep[@$z$ = 0.078;][]{Hu2021}, GRB 200826A/SN \citep[@$z$ = 0.748;][]{Ahumada2021, Rossi2022}. For the events with observations only in the SDSS filters, the $i$-band data is adopted in place of the $R$-band. The light curves plotted in Figure~\ref{XRT} are also corrected for the cosmological expansion. 

The catalogue XRT results for GRB~171010A\footnote{\url{https://www.swift.ac.uk/xrt_live_cat/00020778/}} suggests that the count rate XRT light curve could be best fitted using power-law model with two breaks with following temporal parameters: $\alpha_{1,2,3}$= 2.17$^{+0.29}_{-0.33}$, 0.28$^{+0.91}_{-1.78}$, 1.83$^{+0.16}_{-0.14}$ and T$_{b1,b2}$= 7$^{+4}_{-3}$ $\times ~10^{4}$s, 1.7$^{+0.9}_{-0.3}$ $\times ~10^{5}$s \footnote{where $\alpha_{1,2,3}$ are temporal indices before the first break, between both the breaks, and after the second break, respectively. T$_{b1,b2}$ corresponds to the first and second break time post-detection.}. We considered that the second break is due to jet break (although achromatic behaviour could not be confirmed due to the unavailability of simultaneous multiwavelength data) as the temporal index post-break is consistent with the expected temporal index during jet phase \citep{Chand2019}. On the other hand, the XRT count rate light curve of GRB 171205A\footnote{\url{https://www.swift.ac.uk/xrt_live_cat/00794972/}} could be best fitted using power-law model with three breaks with following temporal parameters: $\alpha_{1,2,3,4}$= 1.26$^{+0.39}_{-0.27}$, 2.30$^{+0.05}_{-0.05}$, 0.02$^{+0.08}_{-0.10}$, 0.98$^{+0.08}_{-0.07}$ and T$_{b1,b2,b3}$= 196$^{+38}_{-9}$s, 5848$^{+641}_{-267}$s, 9.6$^{+1.8}_{-1.4}$ $\times ~10^{4}$s, clearly showing a plateau in the XRT light curve. We noticed that the $X$-ray and $R$-band light curves of nearby GRB~171205A/SN 2017iuk are one of the faintest in the present sample of a total of 15 GRB/SNe, typically expected in case of llGRBs or bursts observed off-axis \citep{DElia2018}. On the other hand, at both wavebands, GRB~171010A/SN~2017htp appeared as one of the brightest GRB/SNe cases, possibly observed on-axis at moderate distances \citep{Chand2019}.

\subsection{Optical light curves modelling}

In this section, we describe methods used to generate the bolometric light curves of SN~2017htp and SN~2017iuk and their semi-analytical modelling using the {\tt MINIM} code \citep{Chatzopoulos2013}.

\subsubsection{Bolometric light curves}
\label{sec:bolometric}

\begin{figure*}[ht!]
\includegraphics[angle=0,scale=0.8]{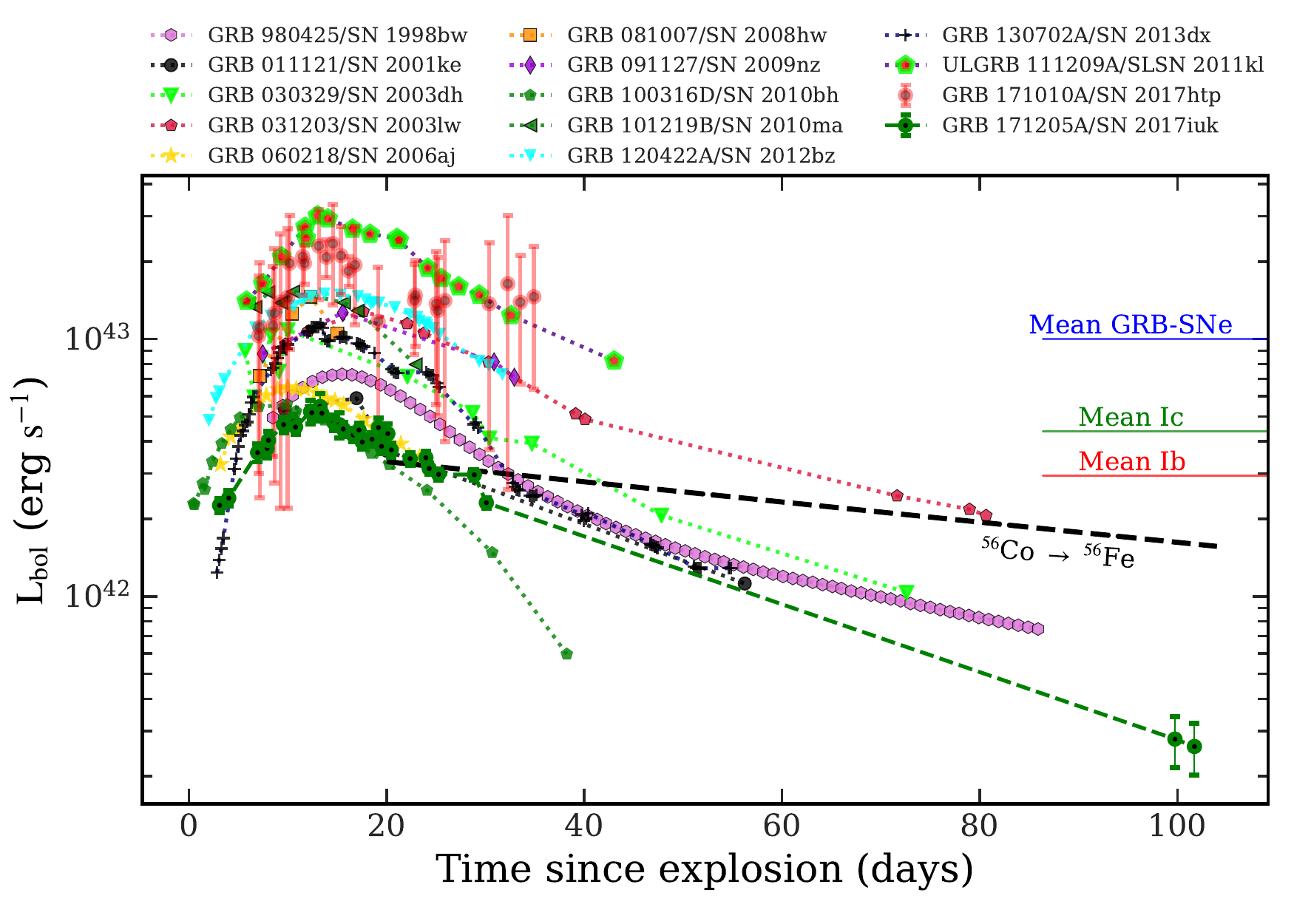}
\centering
\caption{The bolometric light curves of SN~2017htp and SN~2017iuk generated using the {\tt Superbol} code \citep{Nicholl2018} are compared with those of other well-studied GRB-SNe taken from \cite[][and references therein]{Cano2017}. The mean $L_{peak}$ of Type Ib, Ic \citep{Lyman2016}, and GRB-SNe (excluding SLSN 2011kl; \citealt{Cano2017}) along with the $^{56}$Co $\rightarrow$ $^{56}$Fe theoretical decay curve have also been marked for the comparison. SN~2017htp appears to be one of the brightest, and SN 2017iuk is among the faintest GRB-SNe. GRB-SNe presented here exhibit more than an order of magnitude range in bolometric luminosity units.}
\label{fig:bolo_GRB171205A}
\end{figure*}

The pseudo-bolometric light-curves of SN~2017htp ($griz$) and SN~2017iuk ($UgBrViRIzJHK$) are produced using the python-based {\tt Superbol} code \citep{Nicholl2018}, as discussed in \cite{Kumar2020,Kumar2021, Pandey2021}. To compute the multi-band fluxes at similar epochs, interpolations/extrapolations were done wherever necessary by assuming constant colours. Possible contributions of $UV$ and $NIR$ fluxes are included by extrapolating the SED and integrating the observed fluxes to obtain the full bolometric light curves for both events. The full bolometric light-curve ($UV$ to $NIR$; $L_{bol}$) of SN~2017htp is obtained from $\sim$7 to 35 rest-frame days post burst, which exhibits a peak bolometric luminosity ($L_{peak}$) of $\sim$(2.1 $\pm$ 0.9) $\times$ 10$^{43}$ erg s$^{-1}$ (in faded red color; see Figure~\ref{fig:bolo_GRB171205A}). Although, higher error bars are associated with the bolometric light curve of SN~2017htp due to inadequate observed frequency coverage, giving rise to significant uncertainty in the SED fitting. On the other hand, bolometric light-curve of SN~2017iuk (from $\sim$3 to 102 rest-frame days post burst) shows a comparatively lower $L_{peak}$ of $\sim$(0.52 $\pm$ 0.06) $\times$ 10$^{43}$ erg s$^{-1}$ (plotted with the green color in Figure~\ref{fig:bolo_GRB171205A}).

\begin{figure*}[ht!]
\includegraphics[angle=0,scale=0.5]{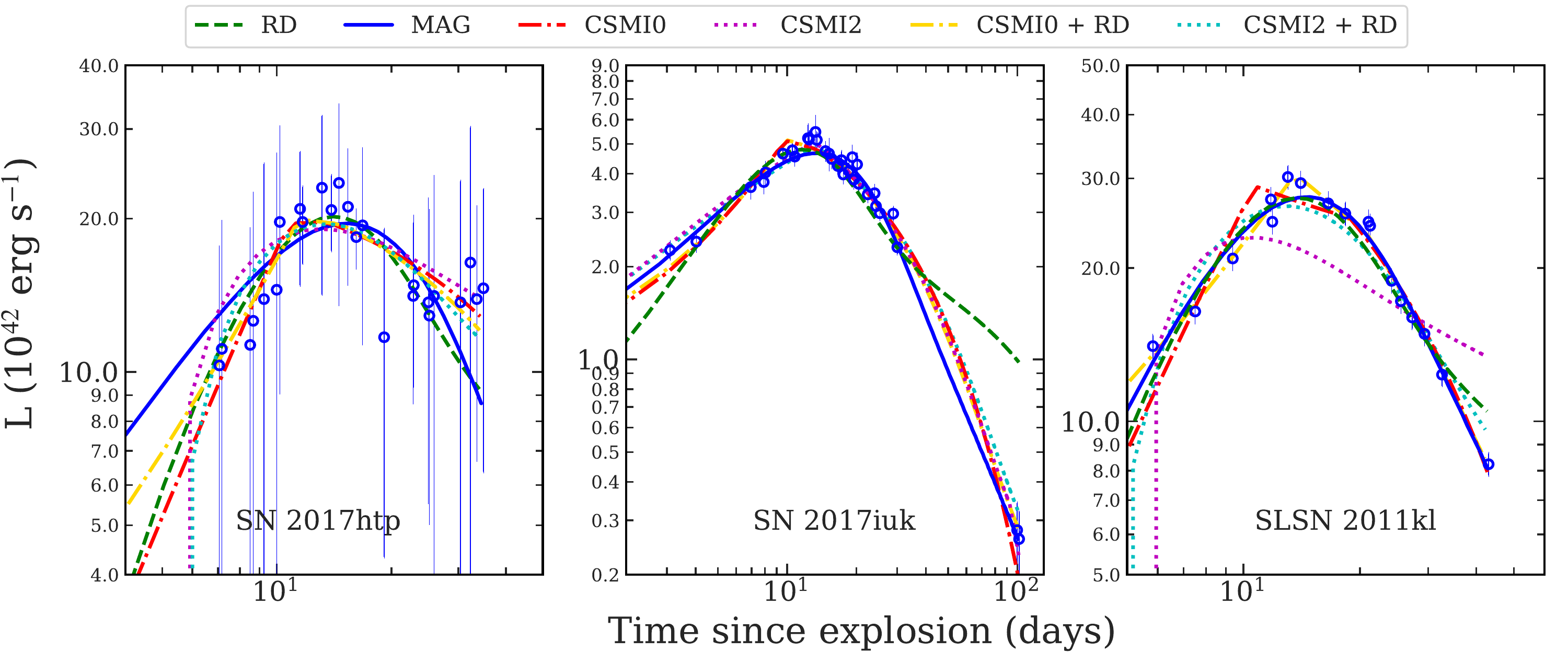}
\caption{Semi-analytical light-curve modelling of SN~2017htp, SN~2017iuk, and SLSN~2011kl is performed using the {\tt MINIM} code \citep{Chatzopoulos2013} and shown in the left, middle, and right panels, respectively. For all three above-mentioned GRB-SNe, $\kappa$ = 0.07 cm$^2$ g$^{-1}$ is adopted.}
\label{fig:minim}
\end{figure*}

The bolometric light curves of SN~2017htp and SN~2017iuk are compared with those of other well-studied GRB-SNe taken from \cite{Cano2017} (see Figure~\ref{fig:LC_magnitudes}). The mean $L_{peak}$ of Type Ib, Ic \citep{Lyman2016}, and GRB-SNe \citep[excluding SLSN 2011kl;][]{Cano2017} are also marked for comparison. SN~2017htp shares one of the brightest $L_{peak}$ than those of other GRB-SNe presented in this study (except SLSN 2011kl; $L_{peak}$ $\sim$[2.9 $\pm$ 0.1] $\times$ 10$^{43}$ erg s$^{-1}$) and also higher than mean $L_{peak}$ quoted for GRB-SNe. On the other hand, SN~2017iuk is one of the low-luminosity GRB-SNe and exhibits a fainter $L_{peak}$ in comparison to those of SN~2017htp, SLSN~2011kl, other plotted GRB-SNe, and mean $L_{peak}$ marked for GRB-SNe, whereas comparable with the mean $L_{peak}$ of SNe~Ic. SLSN~2011kl is the only SLSN associated with a ulGRB and shows the brightest $L_{peak}$ among all GRB-SNe (see Figure~\ref{fig:bolo_GRB171205A}); however, it is the faintest to date in comparison to all other SLSNe~I \citep{Kann2019}.

\begin{table*}[t]
\scriptsize
\caption{Best-fitting parameters for the RD, MAG, CSMI0, CSMI2, CSMI0+RD, and CSMI2+RD models that fitted to the bolometric light curves of SN~2017htp, SN~2017iuk, and SLSN~2011kl using the {\tt MINIM} code \citep{Chatzopoulos2013}. Based on the physical reliability of the estimated parameters, the bolometric light curve of SN~2017htp can be explained using the RD/MAG/CSM models, SLSN~2011kl by the MAG/CSM models, and the light curve of SN~2017iuk can be explained only using the MAG model.}
\begin{center}
\addtolength{\tabcolsep}{-3pt}
\begin{tabular}{c c c c c c c c c c c c c c c}
\hline \hline
 &  &  &  &  \textbf{RD model} & & & & & & \\
SN    & $A_{\gamma}^a$ & $M_{\rm Ni}^b$ & $t_d^c$ & $M_{\rm ej}$  & & & & & $\chi^2$/DOF &\\
$ $ &   (10$^9$) & ($M_\odot$) & (d) & ($M_\odot$) &  & & &  & & \\
    
SN2017htp & 7.25 $\pm$ 2.88 & 0.76 $\pm$ 0.07 & 18.72 $\pm$ 1.64 & 3.27 $\pm$ 0.57 &  &  & &   & 0.11 & \\
SN2017iuk & 3.66 $\pm$ 2.40 & 0.18 $\pm$ 0.01 & 10.76 $\pm$ 0.62 & 2.31 $\pm$ 0.27 &  &  & &  & 11.13 & \\
SLSN2011kl  & 3.51 $\pm$ 2.44 & 1.04 $\pm$ 0.01 & 10.81 $\pm$ 0.16 &  1.56 $\pm$ 0.05 &  &  & &  & 3.76 & \\
    \hline
    
    &         &  & & \textbf{MAG model} & & & & & & \\

    SN    & $R_0^d$ & $E_p^e$ & $t_d$ & $t_p^f$ & $v_{\rm exp}$ & $P_i$ & $B$ &  $M_{\rm ej}$ & $\chi^2$/DOF & \\
    $ $ & ($10^{13}$ cm) & ($10^{51}$ erg) & (d) & (d) & ($10^3$ km s$^{-1}$) & (ms) & ($10^{14}$ G) &  ($M_\odot$) & &\\

SN2017htp & 0.03 $\pm$ 0.03 & 0.12 $\pm$ 0.01 & 17.05 $\pm$ 0.82 & 10.41 $\pm$ 0.92 & 14.47 $\pm$ 0.98 & 12.89 $\pm$ 0.04 & 8.72 $\pm$ 0.26 & 2.80 $\pm$ 0.27 & 0.17 &\\
SN2017iuk & 0.04 $\pm$ 0.02 & 0.03 $\pm$ 0.0001 & 17.06 $\pm$ 0.10 & 9.85 $\pm$ 0.46 & 29.02 $\pm$ 3.51 & 26.42 $\pm$ 0.01 & 18.36 $\pm$ 0.30 & 5.62 $\pm$ 0.07 & 1.04 &\\
SLSN2011kl & 0.03 $\pm$ 0.02 & 0.14 $\pm$ 0.001 & 12.70 $\pm$ 0.29 & 16.08 $\pm$ 0.74 & 29.65 $\pm$ 5.34 & 11.93 $\pm$ 0.01 & 6.49 $\pm$ 0.08 & 3.18 $\pm$ 0.14 & 1.21 &\\
    \hline
    
       &        &  & & \textbf{CSMI0 model} & & & & & & \\
    
    SN & $R_p^g$ & $M_{\rm ej}$ & $M_{\rm csm}^h$ & $\dot{M}$ & $M_{Ni}$& $v_{\rm exp}$ & & & $\chi^2$/DOF &\\
     $ $ & ($10^{13}$ cm) & ($M_\odot$) & ($M_\odot$) & (10$^{-2}$ $\times$ $M_\odot$ yr$^{-1}$) & ($M_\odot$) & ($10^3$ km s$^{-1}$) & & &\\

SN2017htp & 0.08 $\pm$ 0.01 & 5.31 $\pm$ 1.02 & 2.62 $\pm$ 1.24 & 0.0018 $\pm$ 0.0016 & 0.0 $\pm$ 0.0 & 14.15 $\pm$ 1.26 & & & 0.08 &\\
SN2017iuk & 0.09 $\pm$ 0.01 & 2.40 $\pm$ 0.11 & 1.23 $\pm$ 0.02 & 0.0019 $\pm$ 0.0001 & 0.0  $\pm$ 0.0 & 8.42 $\pm$ 0.05 & & & 1.16 &\\
SLSN2011kl & 0.11 $\pm$ 0.02 & 5.61 $\pm$ 0.17 & 1.35 $\pm$ 0.11 & 0.001 $\pm$ 0.00003 & 0.0 $\pm$ 0.0 & 15.27 $\pm$ 0.09 & & & 2.85 &\\
    &          &  & & \textbf{CSMI2 model} & & & & & & \\

SN2017htp & 0.07 $\pm$ 0.03 & 6.46 $\pm$ 1.62 & 4.37 $\pm$ 2.37 & 4.16 $\pm$ 1.08 & 0.0 $\pm$ 0.0 & 14.16 $\pm$ 0.50 & & & 0.13 &\\
SN2017iuk & 1.47 $\pm$ 1.38 & 20.39 $\pm$ 0.87 & 0.98 $\pm$ 0.02 & 19.98 $\pm$ 0.04 & 0.0 $\pm$ 0.0 & 6.55 $\pm$ 0.02 & & & 0.79 &\\
SLSN2011kl & 1.32 $\pm$ 1.26 & 19.88 $\pm$ 0.09 & 1 $\pm$ 0.01 & 20 $\pm$ 0.04 & 0.0 $\pm$ 0.0 & 6.53 $\pm$ 0.02 & & & 0.87 &\\
    &          &  & & \textbf{CSMI0+RD model} & & & & & & \\

SN2017htp & 0.09 $\pm$ 0.02 & 4.53 $\pm$ 1.21 & 2.37 $\pm$ 0.83 & 0.0012  $\pm$ 0.0015 & 0.14 $\pm$ 0.18 & 13.28 $\pm$ 1.34 & & & 0.08 &\\  
SN2017iuk & 0.09 $\pm$ 0.01 & 2.53 $\pm$ 0.14 & 1.03 $\pm$ 0.02 & 0.0013  $\pm$ 0.0001 & 0.02 $\pm$ 0.002 & 8.20 $\pm$ 0.07 & & & 0.96 &\\
SLSN2011kl & 0.08 $\pm$ 0.01 & 3.39 $\pm$ 0.15 & 1.26 $\pm$ 0.07 & 0.00017  $\pm$ 0.00004 & 0.37 $\pm$ 0.03 & 14.45 $\pm$ 0.07 & & & 0.52 &\\
   \\
    &          &  & & \textbf{CSMI2+RD model} & & & & & & \\    

SN2017htp & 7.52 $\pm$ 7.47 & 2.69 $\pm$ 2.39 & 10.75 $\pm$ 4.99 & 3.64 $\pm$ 0.65 & 0.26 $\pm$ 0.08 & 12.67 $\pm$ 1.35 & & & 0.10 &\\
SN2017iuk & 0.04 $\pm$ 0.01 & 19.6 $\pm$ 5.97 & 1.00 $\pm$ 0.01 & 0.20 $\pm$ 0.01 & 0.006 $\pm$ 0.01 & 5.08 $\pm$ 0.14 & & & 0.95 &\\
SLSN2011kl & 0.03 $\pm$ 0.01 & 7.53 $\pm$ 0.36 & 21.57 $\pm$ 4.23 & 1.14 $\pm$ 0.08 & 0.72 $\pm$ 0.01 & 5.11 $\pm$ 0.05 & & & 3.85 &\\
\hline
\end{tabular}
\begin{tablenotes}[para]
        \item[a] $A_\gamma$: optical depth for the gamma-rays measured after 10d post burst.
        \item[b] $M_{\rm Ni}$: radioactive $^{56}$Ni synthesized mass.
        \item[c] $t_d$: effective diffusion timescale.
        \item[d] $R_0$: progenitor radius.
        \item[e] $E_p$: rotational energy.
        \item[f] $t_p$: spin-down timescale of the magnetar.
        \item[g] $R_p$: progenitor radius before the explosion.
        \item[h] $M_{\rm csm}$: CSM mass.
\end{tablenotes}

\end{center}
\label{tab:minim}
\end{table*}

\begin{table*}[t]
\caption{Best fit models and estimated crucial parameters that explain the bolometric light curves of SN 2017htp, SN 2017iuk, and SLSN 2011kl are summarised.}
\addtolength{\tabcolsep}{2pt}
\begin{tabular}{c c c c c c c c}
\hline
SN & Type & $M_{ej}$ & $v_{exp}$  & $P_i$ & $B$ & $R_0$ & Best-fit \\
   &      & ($M_{\odot}$) & ($10^3$ km s$^{-1}$)  & (ms) & ($10^{14}$ G) & ($10^{13}$ cm) & model \\
\hline
\hline
SN 2017htp &  Ic BL  & 2.5-6.3  & 13.5-15.4 & 12.9 $\pm$ 0.1 & 8.7 $\pm$ 0.3 & 0.01--0.11 & RD/MAG/CSM \\ 
SN 2017iuk &  Ic BL  & 5.6 $\pm$ 0.1  & 29.0 $\pm$ 3.5 & 26.4 $\pm$ 0.1 & 18.4 $\pm$ 0.3 & 0.04 $\pm$ 0.02 & MAG\\ 
SLSN 2011kl  &  SLSN I & 3-5.8  & 14.4-35 & 11.9 $\pm$ 0.1 & 6.5 $\pm$ 0.1 &  0.01--0.13 & MAG/CSM\\
\hline
\end{tabular}\label{tab:parameters}
\end{table*}

\subsubsection{Semi-analytical light-curve modelling}
\label{sec:minim}

As a part of the present work, the bolometric light curves of SN~2017htp and SN~2017iuk are used to fit with six semi-analytical models utilising the {\tt MINIM} code: radioactive decay of $^{56}$Ni (RD), spin-down millisecond magnetar (MAG), constant density circumstellar interaction (CSMI0), wind-like CSMI (CSMI2), and HYBRID models \citep[CSMI0+RD and CSMI2+RD;][]{Chatzopoulos2013}; see Figure~\ref{fig:minim}. {\tt MINIM} is a $\chi^2$-minimization code implemented in {\tt C++} and uses the random search technique to find the global minim. Details about the {\tt MINIM} code and the models used here are well described in \cite{Chatzopoulos2012,Chatzopoulos2013}. In terms of light-curve modelling, previous studies attempted to regenerate the light curves of SN~2017htp \citep{Melandri2019} and SN~2017iuk \citep{Izzo2019} only using the RD model. Light-curve modelling of SLSN~2011kl is also discussed in the present analysis; because it is the only known SLSN stretching the limits of brightness and types of SNe connected to GRBs. For all the {\tt MINIM} models discussed above, we adopted the electron-scattering opacity ($\kappa$) = 0.07 cm$^2$ g$^{-1}$, as also suggested by \cite{Izzo2019, Melandri2019}. In the case of RD and MAG models, the M$_{ej}$ values are calculated using equation 10 of \cite{Chatzopoulos2012}. In MAG model, initial spin period of magnetar in ms is estimated using $P_i = (2 \times 10^{50}$erg s$^{-1}$ /$E_{p})^{0.5} \times 10$ and magnetic field is calculated in gauss as $B = (1.3 P_{10}^{2}/t_{p,yr})^{0.5} \times 10^{14}$; here $t_{p,yr}$ stands for the spin-down timescale of the magnetar in years \citep{Chatzopoulos2013}.

In the present analysis, at the same time, multiple models reproduced the bolometric light curves of all GRB-SNe with $\chi^2$/DOF $<$1 (see Table~\ref{tab:minim}). The possible reasons behind $\chi^2$/DOF $<$ 1 are the association of higher error bars to the data points, a lower number of data points, and the larger number of fitting parameters. Hence, in the current analysis, $\chi^2$/DOF is not used as a benchmark to select the most suitable model but is only used to choose the model parameters that adequately regenerate the light curve. At the same time, we focused more on the level of the physical significance of the parameters estimated by different models, as also suggested by \cite{Kumar2021, Pandey2021}. 

Statistically, the bolometric light curve of SN~2017htp is well-reproduced by all the six models (RD, MAG, CSMI0, CSMI2, CSM0+RD, and CSM2+RD) with $\chi^2$/DOF $<$ 1 (see the left panel of Figure~\ref{fig:minim}). Hence as discussed above, to choose the most reliable model, we concentrated on the physical reliability of the estimated parameters. In the case of CSMI2 and CSMI2+RD models, the values of progenitor mass-loss rates ($\dot{M}$) are unphysically higher; therefore, these can not be considered as most suitable models for SN~2017htp. On the other hand, the parameters estimated by the other four models (RD, MAG, CSMI0, CSMI0+RD) are physically viable. So, RD, MAG, and CSMI models are equally favourable in explaining the light curve of SN~2017htp and based on the light curve modelling results only; it is not possible to discriminate any of these models. However, along with the light curve modelling results, the prompt analysis also supports a magnetar as a central engine powering source for SN~2017htp (see Section~\ref{sec:central_engine}). All the parameters derived using the {\tt MINIM} modelling of SN~2017htp are listed in Table~\ref{tab:minim}.

In the case of SN~2017iuk, the RD model is unable to reproduce the late time observations (at $\sim$100d) and exhibits a poorer fit with $\chi^2$/DOF = 11.13. On the other hand, the rest five models reproduced the bolometric light curve of SN~2017iuk with $\chi^2$/DOF $\approx$1 (see the middle panel of Figure~\ref{fig:minim}). However, CSMI0, CSMI2, CSMI0+RD, and CSMI2+RD models gave unphysical values of expansion velocities $v_{exp}$ $<$8,500 km s$^{-1}$, very low for a type Ic-BL SN and also much lower than the value estimated by \cite{Wang2018, Izzo2019} based on the spectral analysis. In addition, the values of $\dot{M}$ constrained by the CSMI2 and CSMI2+RD models are also unphysically higher. So, RD, CSMI, and RD+CSMI can not be considered probable powering mechanisms for SN 2017iuk. On the other hand, the bolometric light curve of SN~2017iuk is well-reproduced by the MAG model with physically viable parameters. So, the MAG model seems like the most suitable one for SN~2017iuk, which is also in agreement with the results discussed in Section~\ref{sec:central_engine}. All the fitted parameters for the SN~2017iuk using the {\tt MINIM} code are tabulated in Table~\ref{tab:minim}.

In the case of SLSN~2011kl light curve modelling (see the right panel of Figure~\ref{fig:minim}), the $M_{ej}$ value derived using the RD model is very low in comparison to those generally observed for SLSNe~I \citep{Nicholl2015}. At the same time, CSMI2 and CSMI2+RD models suggested comparatively lower values of $v_{exp}$ and higher values of $\dot{M}$ in comparison to what is generally seen for GRB-SNe \citep{Woosley2006,Woosley2006a, Smith2014}. On the other hand, MAG, CSMI0, and CSMI0+RD models reproduced the bolometric light curve of SLSN~2011kl with physically reliable parameters and $\chi^2$/DOF close to one. The complete list of estimated parameters using the above models for SLSN~2011kl is tabulated in Table~\ref{tab:minim}. So, based on the present analysis, CSMI and spin down millisecond magnetar are the possible powering sources of SLSN~2011kl.

In all, results from the light-curve modelling using the {\tt MINIM} code clearly indicate that the model considering a spin down millisecond magnetar as a central engine powering source is the only model which can explain the light curves of all three GRB-SNe discussed presently. In addition, the light-curve model fittings help to constrain various crucial parameters of SN~2017htp, SN~2017iuk, and SLSN~2011kl that are tabulated in Table~\ref{tab:parameters}.

\begin{figure}[ht!]
\centering
\includegraphics[angle=0,scale=0.8]{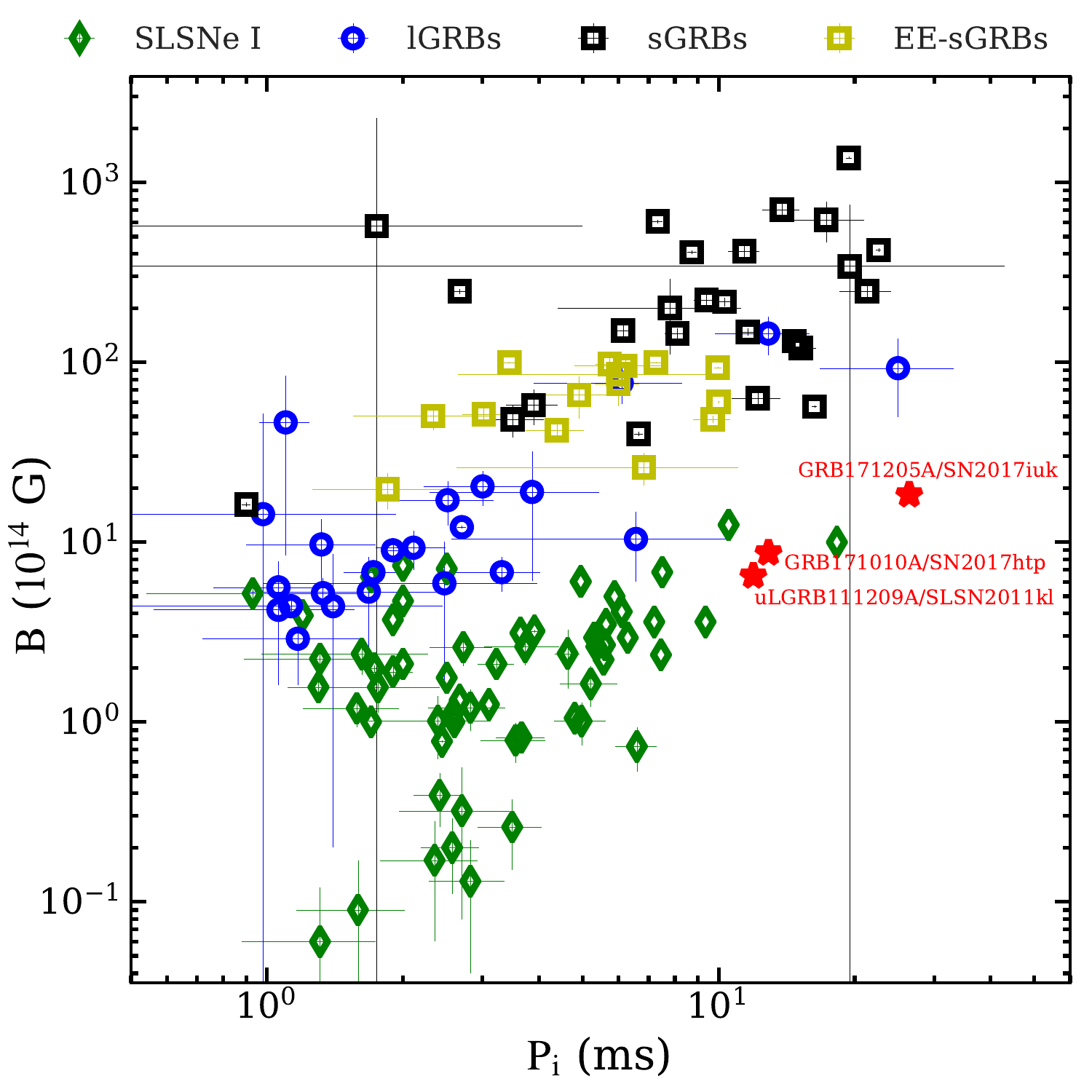}
\caption{Modelled values of $B$ versus $P_i$ as constrained for SN~2017htp, SN~2017iuk, and SLSN~2011kl are compared with those of well studied sGRBs \citep{Suvorov2021}, EE-sGRBs \citep{Gibson2017}, lGRBs \citep{Liang2018}, and SLSNe~I \citep[][and references therein]{Kumar2021}.}
\label{fig:Pi_B}
\end{figure}

\subsection{Magnetars and $B$ versus $Pi$}
Magnetars are expected to be the central powering sources for many energetic transients, including a certain class of GRBs and SNe \citep{Duncan1992, Usov1992, Wheeler2000, Metzger2015, Lin2020ApJ}. Magnetars can produce lGRBs by inducing the relativistic Poynting-flux or magnetically driven baryonic jets \citep{Bucciantini2008}. At the same time, a strong magnetic field of the magnetar can dissipate its rotational energy to energise the associated SN through magnetic braking or magnetic dipole radiation \citep{Duncan1992, Usov1992}. Overall, magnetars as the central engine powering sources can explain the different types of light-curve properties with diverse sets of physical parameters like $B$ (10$^{14}$--10$^{15}$ gauss) and $P_i$ (a few ms), see \cite{Duncan1992}. Hence, $B$ and $P_i$ values of a newborn magnetar can govern the energy of the resulting GRB and SN, e.g., a fast-rotating magnetar may produce a more energetic GRB \citep{Zou2019, Zou2021}. The difference in properties of magnetars originated from the core-collapse of a massive star and the merger of compact objects are presented by \cite{Zou2021}. They found that the magnetars associated with the core-collapse of massive stars exhibit weaker $B$ and a shorter $P_i$, nearly one to two orders of magnitude compared to those generated from the binary compact mergers. Although, \cite{Cano2016} shows that a magnetar-based central engine cannot be solely responsible for producing the observed luminosities of GRB/SNe.

In the present analysis, the $B$ versus $P_i$ estimated for SN~2017htp, SN~2017iuk, and SLSN~2011kl using the MAG model under the {\tt MINIM} code are compared with those reported for a sample of well-studied sGRBs \citep{Suvorov2021}, Extended Emission-sGRBs \citep[EE-sGRBs;][]{Gibson2017}, lGRBs \citep{Liang2018}, and SLSNe~I \citep{Kumar2021}; see Figure~\ref{fig:Pi_B}. Though, we caution that the compiled values of $B$ and $P_i$ have been adopted from diverse sources using different methods to estimate the parameters, so any possible underlying systematics/biasing can not be ruled out. For example, in the case of sGRBs, \cite{Suvorov2021} used 25 Swift-XRT observed $X$-ray light curves of bursts with a plateau and fitted to luminosity profiles suitable for precessing oblique rotators. \cite{Gibson2017} used afterglows of 15 EE-sGRBs obtained from the Swift archive and fitted with the model developed by combining fallback accretion into the magnetar propeller model. For lGRBs, \cite{Liang2018} analysed Swift-XRT light curves of 101 events with plateau phases and investigated the possibility of a fast-rotating black hole and a rapidly spinning magnetar as a central engine powering source. The $B$ and $P_i$ values for SLSNe~I are taken from figure 10 of \cite{Kumar2021}, where the authors adopted complied values from previous studies.

Based on the sample comparison discussed above, most of the sGRBs seem to have higher $P_i$ and $B$ values than those for lGRBs and SLSNe~I  (see Figure~\ref{fig:Pi_B}), as recently also suggested by \cite{Zou2021}. However, most of the EE-sGRBs appear to have $B$ values in between those of sGRBs and lGRBs. On the other hand, SLSNe~I exhibit a broad range of $P_i$ but comparatively lower values of $B$. Among the present limited sample of three GRB-SNe, SN~2017htp, SN~2017iuk, and SLSN~2011kl appear to occupy a different space in the $B$-$P_i$ diagram with $B$ values closer to lGRBs but somewhat higher values of $P_i$ compared to most of the lGRBs and SLSNe~I; however, lower than those estimated in case sGRBs. Although, investigation with a larger sample of GRB-SNe is a must.

\section{Summary and Conclusions}
\label{sec:conclusion}

In this paper, we present the prompt emission and late-time optical observations and semi-analytical light curve modelling of two interesting and nearby GRBs associated with SNe of extreme brightness (GRB~171010A/SN~2017htp and GRB~171205A/SN~2017iuk). We noticed that the prompt emission properties such as spectral hardness, \tninty, and \mvts are comparable for both the bursts. However, GRB~171010A (third GBM fluent burst) is significantly luminous in comparison to GRB~171205A. In fact, GRB~171205A belongs to the low-luminous family of GRBs and is an outlier to the well-known Amati correlation for lGRBs. The time-integrated spectra of GRB~171010A demand for a thermal component along with typically observed non-thermal \sw{Band} function, suggesting a hybrid jet composition. We notice that $E_{\rm \gamma, beamed}$ values for GRB~171010A and GRB~171205A are less than the maximum possible rotational energy budget of a typical magnetar, indicating a central engine powering source for both the bursts. Additionally, during the plateau phase of GRB~171205A, the $E_{\rm X, iso}$ is the lowest and $E_{\rm K, iso}$ is the second-lowest in comparison to the sample studied by \cite{Liang2018}; whereas both the values of kinetic energies are lower than the maximum energy budget of magnetars, which also favours a magnetar based powering source for GRB~171205A.

Well-calibrated late-time optical observations of GRB~171010A/SN~2017htp ($\sim$43d post burst) and GRB~171205A/\\SN~2017iuk ($\sim$105d post burst) acquired using the 4K$\times$4K CCD Imager at the 3.6m DOT are not only valuable for constraining the GRB-SNe properties but also provide the longest temporal coverage to deeper limits for GRB~171205A/SN~2017iuk. The late-time optical observations also demonstrate the imaging capabilities of the 3.6m DOT for such interesting transients up to faint limits. The new multi-band optical observations, along with the published ones, helped to generate the bolometric light curves of SN~2017htp and SN~2017iuk. The bolometric light-curve of SN~2017htp exhibited peak-luminosity of $\sim$(2.1 $\pm$ 0.9) $\times$ 10$^{43}$ erg s$^{-1}$, comparable to most luminous GRB-SNe. On the other hand, SN~2017iuk is one of the low-luminous GRB-SNe with a peak luminosity of $\sim$(0.52 $\pm$ 0.06) $\times$ 10$^{43}$ erg s$^{-1}$, lower than SN~2017htp, SLSN~2011kl, and other GRB-SNe discussed in this study. SLSN~2011kl is the only GRB associated SLSN with the highest peak luminosity of $\sim$(2.9 $\pm$ 0.1) $\times$ 10$^{43}$ erg s$^{-1}$, luminous by a factor of around five than SN~2017iuk. 

The semi-analytical light-curve modelling on the bolometric light curves of SN~2017htp, SN~2017iuk, and SLSN~2011kl was performed using the {\tt MINIM} code. The modelling outputs for SN~2017htp cannot discriminate among the three models (RD, MAG, and CSMI) and make reasonable light-curve fits for all three. On the other hand, the bolometric light curve of SN~2017iuk can be reproduced using the MAG model only by giving good fits and viable physical parameters. However, the bolometric light curve of SLSN~2011kl can be regenerated using both the MAG and CSMI models. In all, among three models representing different powering mechanisms, the spin-down millisecond magnetar emerged as the common powering source under the MAG model explaining the bolometric light curves of all three GRB-SNe discussed here. In addition, the light curve modelling results helped to constrain many crucial parameters of SN~2017htp, SN~2017iuk, and SLSN~2011kl. Amid three GRB-SNe discussed presently, expansion velocity, spin-down time-scale, and magnetic field values of SN~2017iuk are found higher in compassion to those of SN~2017htp and SLSN~2011kl; however, there is no significant difference in the ejecta masses. 

We also compared the $B$ and $P_i$ values of SN~2017htp, SN~2017iuk, and SLSN~2011kl to a sample of transients having evidence of magnetar origin (from diverse sets of studies), which includes sGRBs, EE-sGRBs, lGRBs, and SLSNe~I. All three GRB-SNe discussed presently appear to have $B$ values higher than SLSNe~I, closer to lGRBs and lower than those of sGRBs. Whereas $P_i$ values of GRB-SNe are higher than those of SLSNe~I and lGRBs, but lower than sGRBs. Even the SLSN~2011kl connected with ulGRB 111209A exhibits higher values of $B$ and $P_i$ than those observed for other SLSNe~I. The present comparison of $B$ and $P_i$ for diverse sets of transients demanding magnetars indicate a continuum among possible progenitors and the powering source, giving rise to immense energies across the electromagnetic spectrum. In the near future, late-time observations of many such transients with the 3.6m DOT and other facilities will be beneficial in understanding the underlying possible powering mechanisms constraining physical parameters in more detail.

\section*{Acknowledgements}
This study uses the data taken from the 4K$\times$4K CCD Imager at the 3.6m Devasthal Optical telescope (3.6m DOT), and SBP is grateful to the staff members of the 3.6m DOT for their consistent support and help during observations. AK and SBP acknowledge the discussions related to the light-curve modelling results with Prof. Jozsef Vink{\'o}. AK is also thankful to Dr. Zach Cano for the scientific discussion and for sharing the ASCII files. SBP, RG, and AA acknowledge BRICS grant DST/IMRCD/BRICS/Pilotcall/ProFCheap/2017(G) and DST/JSPS grant DST/INT/JSPS/P/281/2018 for this work. SBP and RG acknowledge the financial support of ISRO under AstroSat archival Data utilization program (DS$\_$2B-13013(2)/1/2021-Sec.2). AA also acknowledges funds and assistance provided by the Council of Scientific \& Industrial Research (CSIR), India, with file no. 09/948(0003)/2020-EMR-I. This work has utilised the NED, operated by the Jet Propulsion Laboratory, California Institute of Technology, under contract with NASA. We acknowledge the use of NASA’s Astrophysics Data System Bibliographic Services.

\newpage
\appendix
\section*{Appendix}
\setcounter{table}{0}
\renewcommand{\thetable}{A\arabic{table}}

\begin{table*}[!h]
\addtolength{\tabcolsep}{14pt}
\caption{Calibrated magnitudes of the secondary standard stars in the field of GRB~171010A/SN 2017htp observed using the 4K$\times$4K CCD Imager mounted at the axial port of the 3.6m DOT \citep{Pandey2018, Kumar2022}, shown in the left panel of Figure~\ref{fig:finding_171010A} (S1--S9).}
\begin{tabular}{lccccc}
\hline
\textbf{ID}&\textbf{U}&\textbf{B}&\textbf{V}&\textbf{R}&\textbf{I}\\
\hline 
\hline	
  1 &  19.08$\pm$0.06 &  19.19$\pm$0.03 &  18.42$\pm$0.02 &  17.86$\pm$0.06 &  17.32$\pm$0.04 \\
  2 &  17.35$\pm$0.05 &  17.36$\pm$0.03 &  16.57$\pm$0.02 &  16.04$\pm$0.06 &  15.48$\pm$0.04 \\
  3 &  18.40$\pm$0.05 &  18.22$\pm$0.03 &  17.47$\pm$0.02 &  17.07$\pm$0.06 &  16.49$\pm$0.04 \\
  4 &  18.80$\pm$0.06 &  18.34$\pm$0.03 &  17.37$\pm$0.02 &  16.73$\pm$0.06 &  16.08$\pm$0.04 \\
  5 &  18.46$\pm$0.05 &  18.21$\pm$0.03 &  17.35$\pm$0.02 &  16.77$\pm$0.06 &  16.19$\pm$0.04 \\
  6 &  17.17$\pm$0.05 &  16.78$\pm$0.03 &  15.82$\pm$0.02 &  15.38$\pm$0.06 &  14.85$\pm$0.04 \\
  7 &  17.09$\pm$0.05 &  17.02$\pm$0.03 &  16.39$\pm$0.02 &  15.80$\pm$0.06 &  15.07$\pm$0.04 \\
  8 &  20.98$\pm$0.12 &  19.51$\pm$0.03 &  17.85$\pm$0.02 &  16.68$\pm$0.06 &  15.51$\pm$0.04 \\
  9 &  20.72$\pm$0.10 &  19.41$\pm$0.03 &  17.75$\pm$0.02 &  16.60$\pm$0.06 &  15.56$\pm$0.04 \\
\hline
\label{tab:171010A_secondary_stars}
\end{tabular}
\end{table*}

\begin{table*}[!h]
\addtolength{\tabcolsep}{14pt}
\caption{Calibrated magnitudes of the secondary standard stars in the field of GRB 171205A/SN 2017iuk observed using the 4K$\times$4K CCD Imager plus 3.6m DOT, shown in the upper-right panel of Figure~\ref{fig:finding_171205A} (S1--S7).}
\begin{tabular}{lccccc}
\hline
\textbf{ID}&\textbf{U}&\textbf{B}&\textbf{V}&\textbf{R}&\textbf{I}\\
\hline 
\hline				
  1 &  18.60$\pm$0.21 &  18.50$\pm$0.07 &  17.54$\pm$0.03 &  17.06$\pm$0.03 &  16.72$\pm$0.05 \\
  2 &  19.45$\pm$0.24 &  18.63$\pm$0.07 &  17.19$\pm$0.03 &  16.34$\pm$0.03 &  15.46$\pm$0.04 \\
  3 &  17.32$\pm$0.21 &  17.19$\pm$0.07 &  16.30$\pm$0.03 &  15.81$\pm$0.03 &  15.41$\pm$0.04 \\
  4 &  18.12$\pm$0.21 &  17.91$\pm$0.07 &  17.01$\pm$0.03 &  16.52$\pm$0.03 &  16.08$\pm$0.04 \\
  5 &  18.32$\pm$0.21 &  17.86$\pm$0.07 &  16.85$\pm$0.03 &  16.26$\pm$0.03 &  15.69$\pm$0.04 \\
  6 &  19.66$\pm$0.25 &  19.35$\pm$0.08 &  18.03$\pm$0.03 &  17.31$\pm$0.04 &  16.62$\pm$0.05 \\
  7 &  19.77$\pm$0.27 &  19.21$\pm$0.08 &  17.75$\pm$0.03 &  16.89$\pm$0.04 &  15.82$\pm$0.05 \\
\hline
\label{tab:secondary_stars_aa}
\end{tabular}
\end{table*}

\newpage

\bibliographystyle{mnras}
\bibliography{casrefs}


\end{document}